\documentclass[twocolumn,epjc3]{svjour3}
\usepackage[utf8]{inputenc}
\usepackage{amsmath,amsfonts,amssymb,slashed,graphicx,hyperref,color,cite}
\usepackage{multirow,arydshln}
\hypersetup{colorlinks=true,linkcolor=blue,citecolor=magenta,
  filecolor=magenta, urlcolor=cyan}
\journalname{Eur. Phys. J. C}
\graphicspath{{figures/}}

\def\ie{{\it i.e.}}
\def\eg{{\it e.g.}}

\newcommand{\be}{\begin{equation}}
\newcommand{\ee}{\end{equation}}
\def\bsp#1\esp{\begin{split}#1\end{split}}
\def\bpm{\begin{pmatrix}}
\def\epm{\end{pmatrix}}

\def\lag{{\cal L}}
\def\sss{\scriptscriptstyle}
\def\as{\alpha_{\sss s}}
\def\gw{g_{\sss W}}
\def\qL{Q}
\def\uR{u}
\def\dR{d}

\newcommand{\ch}{{\sc CalcHep}}
\newcommand{\del}{{\sc Delphes}~3}
\newcommand{\fa}{{\sc FeynArts}}
\newcommand{\fj}{{\sc FastJet}}
\newcommand{\fr}{{\sc FeynRules}}
\newcommand{\ma}{{\sc MadAnalysis}~5}
\newcommand{\maddm}{{\sc MadDM}}
\newcommand{\mfks}{{\sc MadFKS}}
\newcommand{\msp}{{\sc MadSpin}}
\newcommand{\mstr}{{\sc MadSTR}}
\newcommand{\mwdth}{{\sc MadWidth}}
\newcommand{\mg}{{\sc MG5\_aMC}}
\newcommand{\micromegas}{{\sc MicrOMEGAs}}
\newcommand{\mthmtc}{{\sc Mathematica}}
\newcommand{\nloct}{{\sc NLOCT}}
\newcommand{\py}{{\sc Pythia}~8}
\def\dmsimpt{{\tt DMSimpt}}

\begin{document}

\title{A universal framework for $t$-channel dark matter models}

\author{Chiara Arina\thanksref{e1,addr1}
        \and
        Benjamin Fuks\thanksref{e2,addr2,addr3} 
        \and 
        Luca Mantani\thanksref{e3,addr1}}

\thankstext{e1}{e-mail: chiara.arina@uclouvain.be}
\thankstext{e2}{e-mail: fuks@lpthe.jussieu.fr}
\thankstext{e3}{e-mail: luca.mantani@uclouvain.be}

\institute{Centre for Cosmology, Particle Physics and Phenomenology (CP3), Universit\'e catholique de Louvain, B-1348 Louvain-la-Neuve, Belgium\label{addr1}
\and~Sorbonne Universit\'e, CNRS, Laboratoire de Physique Th\'eorique et  Hautes \'Energies, LPTHE, F-75005 Paris, France\label{addr2}
\and~Institut Universitaire de France, 103 boulevard Saint-Michel, 75005 Paris, France\label{addr3}}

\date{Received: date / Accepted: date}

\maketitle

\begin{abstract}
We present the \dmsimpt\ model implementation in \fr, which aims to offer a
unique general framework allowing for all simulations relevant for simplified
$t$-channel dark matter models at colliders and for the complementary
cosmology calculations. We describe how to match next-to-leading-order QCD
fixed-order calculations with parton showers to derive robust bounds and
predictions in the context of LHC dark matter searches, and moreover validate
two model restrictions (relevant for Dirac and Majorana fermionic dark matter
respectively) to exemplify how to evaluate dark matter observables to constrain the
model parameter space. More importantly, we emphasise how to achieve
these results by using a combination of publicly available automated tools, and
discuss how dark matter predictions are sensitive to the model file
and software setup.
All files, together with illustrative \mthmtc\ notebooks, are available from the
URL \url{http://feynrules.irmp.ucl.ac.be/wiki/DMsimpt}.
% \keywords{Beyond Standard Model \and Dark Matter simplified models \and Dark Matter LHC phenomenology \and Dark matter tools}
% \PACS{95.35.+d \and 95.30.Cq}
\end{abstract}

%%%%%%%%%%%%%%%%%%%%%%%%%%%%%%%%%%%%%%%%%%%%%%%%%%%%%%%%%%%%%%%%%%%%%%%%%%%%%%%
\section{Introduction}\label{intro}

Despite of convincing evidence for its existence~\cite{Bertone:2010zza}, dark
matter still evades direct detection both in dedicated underground nuclear
recoil experiments and at colliders. Getting insights on the nature of dark
matter and the way in which it interacts with the Standard Model particles
therefore consists in one of the hot topics in particle and astroparticle
physics today. One potential strategy that could shed light on this matter
involves simplified models~\cite{Alwall:2008ag,Alves:2011wf} in which the
Standard Model is minimally extended in terms of particles and new couplings.
This approach allows for the exploration of viable dark matter scenarios in a
model-independent way and the comparison of theoretical predictions with results
of direct, indirect and collider searches. This however requires the ability of
making predictions for large classes of models, both at colliders and for what
concerns cosmology.

The \fr\ package~\cite{Alloul:2013bka} offers such a
possibility, as from a unique \fr\ implementation of any given dark matter
model, it is subsequently possible to generate model files suitable for various
high-energy physics tools such as \mg~\cite{Alwall:2014hca},
\maddm~\cite{Ambrogi:2018jqj} or \micromegas~\cite{Belanger:2018mqt}. Following
the general strategy for new physics computations outlined in
ref.~\cite{Christensen:2009jx}, such a joint usage of various packages has two
major advantages in the dark matter context. First, it allows for the
straightforward and automatic calculation of the dark matter relic density,
as well as of the
direct and indirect detection cross sections to verify the cosmological
viability of any model. Second, it enables the extraction of the exclusion
levels of various searches at colliders through the automated generation of
realistic collision events and the recasting
of the corresponding LHC analyses. In the latter case, the \mg\ framework allows
in particular for simulations including next-to-leading order corrections in
$\as$, so that predictions are accurate enough to derive robust constraints
when LHC recasting is at stake through, \eg, the \ma\
platform~\cite{Conte:2018vmg} that includes, from version 1.8, the
propagation of the theoretical uncertainties on the signal predictions
up to the derived exclusion levels~\cite{Araz:2019otb}.

In most simplified models for dark matter, the dark matter is assumed to be a
single massive particle that interacts weakly with the Standard Model through a
mediator particle. In $s$-channel setups~\cite{Fox:2012ru,Haisch:2013ata,%
Backovic:2015soa,%
Abercrombie:2015wmb}, the mediator is colour-singlet and is enforced to couple
to a pair of either dark matter particles, or Standard Model particles.
Such a configuration generally arises in scenarios in which the dark matter
stability is guaranteed by a $\mathbb{Z}_2$ discrete symmetry under which all
Standard Model fields and the mediator are even, and the dark matter particle is
odd. A comprehensive approach for achieving automatic and straightforward
cosmological calculations and collider simulations for $s$-channel dark matter
models has been recently proposed~\cite{Backovic:2015soa}, the cornerstone being
a unique \fr\ implementation driving any subsequent computation.

The present
work is dedicated to a general implementation, in the \fr\ package, of a large
set of
$t$-channel dark matter models in which the mediator interacts with one of the
Standard Model quarks and dark matter.
We have used this \fr\ implementation to generate a UFO library~\cite{%
Degrande:2011ua} that can subsequently be imported in programmes like \mg\ or
\maddm\ for undertaking various simulations and computations for a large class
of $t$-channel dark matter models. Our implementation in
particular allows for collider simulations systematically including
next-to-leading-order (NLO) QCD corrections to all new physics processes
involving either the dark matter particle, the mediator or both.

Such a
possibility requires however a specific treatment of the real emission
contributions that feature, in $t$-channel dark matter models, narrow
$s$-channel resonances. Real-emission corrections to a given
process (\eg~dark matter pair-production whose real emission contributions
include the production of a system comprised of a dark-matter pair and a jet)
may indeed include partonic
sub-processes featuring an $s$-channel resonance corresponding to another Born
process (\eg~mediator/dark matter associated production) followed by a tree-level
decay (\eg~mediator decay into dark matter and jets).
The integration of such contributions over the phase space leads to a
growth proportional to an inverse power of the resonance width, so that such
contributions could be numerically dominant
and apparently spoil the convergence of the perturbative series. 
We recall that this type of configuration also exists in the Standard Model, in
particular in the context of $tW$ production. Real correction to the latter
process include diagrams describing the production of an
$s$-channel resonant $t\bar t$ final state, followed by a top decay into a $Wb$
system.

Moreover, when all new physics processes allowed by the model are considered as
a whole (as each subprocess contributes to the new physics signal), these
resonant diagrams could be double-counted and lead to incorrect
predictions. They therefore need to be treated consistently. Different
strategies to treat these resonances have been
recently automated within the \mg\ framework~\cite{Frixione:2019fxg}, hence
enabling NLO QCD simulations for the considered $t$-channel dark matter models
in a way that is as easy as for the $s$-channel case.

In order to illustrate the strength of our approach, we focus on two limiting
cases and study their phenomenology at colliders and in cosmology, which
allows for the validation of our implementation. We in
particular compare the performances of \maddm\ and \micromegas\ and present,
for the first time, automated computations for loop-induced processes relevant
for dark matter indirect detection. Such a feature, which will be available from
the next release of \maddm, greatly eases the phenomenological analysis of
$t$-channel dark matter models. More specifically, we consider the case of a
fermionic dark matter particle whose interactions with the Standard Model are
mediated by a scalar particle coupling to the right-handed up quark, both for
what concerns Dirac and Majorana dark matter. In the following, we
coin these two configurations, that have been vastly studied in the literature
(as shown \eg\ in refs.~\cite{Hisano:2011cs,Bai:2013iqa,Garny:2013ama,%
Chang:2013oia,An:2013xka,DiFranzo:2013vra,%
Giacchino:2014moa,Garny:2014waa,Garny:2015wea,Ibarra:2015fqa,Berlin:2015njh,%
Hisano:2015bma,Ko:2016zxg,Carpenter:2016thc,ElHedri:2017nny,Hisano:2018bpz}) and
that are particularly promising for LHC and dark matter
searches (see \eg\ refs.~\cite{Papucci:2014iwa,Mohan:2019zrk}), the
{\tt S3D\_uR} and {\tt S3M\_uR} models, respectively.

The rest of the paper is structured as follows. In the next section, we present
the model conventions, its implementation into \fr\ and the restrictions (\ie\
the limiting cases) shipped with our general implementation. In
section~\ref{sec:nlomatch}, we detail how to match NLO QCD calculations with
parton showers for collider simulations, providing extensive details on how to make use of \mg\ in order to ensure a consistent treatment of the
resonant contributions appearing at ${\cal O}(\as)$.
We then present, for the first time, total rate and differential distributions
extracted from accurate predictions matching NLO QCD calculations with parton
showers, and derive the corresponding constraints from selected LHC searches. In
section~\ref{sec:dmsearches}, we briefly outline the dark matter observables
relevant for $t$-channel dark matter models, how to compute them with \maddm,
and present the results for the {\tt S3D\_uR} and {\tt S3M\_uR} model
restrictions to validate our implementation against known results.
We summarise our work in section~\ref{sec:concl}.

%%%%%%%%%%%%%%%%%%%%%%%%%%%%%%%%%%%%%%%%%%%%%%%%%%%%%%%%%%%%%%%%%%%%%%%%%%%%%%%

\section{\fr\ implementation and conventions}\label{sec:model}
\subsection{Generalities}\label{sec:general}

\begin{table*}
\centering
\renewcommand{\arraystretch}{1.4}
\setlength\tabcolsep{8pt}
\begin{tabular}{c c c c c c}
  Field & Spin & Repr. & Self-conj. & \fr\ name & PDG\\
  \hline\hline
  $\tilde S$     & 0   & $({\bf 1}, {\bf 1}, 0)$ & yes & {\tt Xs} & 51\\
  $S$            & 0   & $({\bf 1}, {\bf 1}, 0)$ & no  & {\tt Xc} & 56\\
  $\tilde\chi$   & 1/2 & $({\bf 1}, {\bf 1}, 0)$ & yes & {\tt Xm} & 52\\
  $\chi$         & 1/2 & $({\bf 1}, {\bf 1}, 0)$ & no  & {\tt Xd} & 57\\
  $\tilde V_\mu$ & 1   & $({\bf 1}, {\bf 1}, 0)$ & yes & {\tt Xv} & 53\\
  $V_\mu$        & 1   & $({\bf 1}, {\bf 1}, 0)$ & no  & {\tt Xw} & 58\\
  \hline\hline
  $\varphi_{\sss Q} = \bpm\varphi^{(u)}_{\sss Q}\\ \varphi^{(d)}_{\sss Q}\epm$
     & 0 & $({\bf 3}, {\bf 2},  \frac16)$ & no & ${\tt YS3Q} =
        \bpm {\tt YS3Qu} \\ {\tt YS3Qd}\epm$&
        $\begin{array}{l l l l}
         \varphi^{(u)}_{\sss Q}: & 1000002 & 1000004 & 1000006\\
         \varphi^{(d)}_{\sss Q}: &1000001 & 1000003 & 1000005\end{array}$\\
  $\varphi_{\sss u}$
     & 0 & $({\bf 3}, {\bf 1},  \frac23)$ & no & {\tt YS3u}
     & $\begin{array}{l l l} 2000002 & 2000004 & 2000006\end{array}$\\
  $\varphi_{\sss d}$
     & 0 & $({\bf 3}, {\bf 1}, -\frac13)$ & no & {\tt YS3d}
     & $\begin{array}{l l l} 2000001 & 2000003 & 2000005\end{array}$\\
  \hline
  $\psi_{\sss Q} = \bpm\psi^{(u)}_{\sss Q}\\ \psi^{(d)}_{\sss Q}\epm$
     & 1/2 & $({\bf 3}, {\bf 2},  \frac16)$ & no &${\tt YF3Q} =
        \bpm {\tt YF3Qu} \\ {\tt YF3Qd}\epm$&
        $\begin{array}{l l l l}
         \psi^{(u)}_{\sss Q}: & 5910002 & 5910004 & 5910006\\
         \psi^{(d)}_{\sss Q}: & 5910001 & 5910003 & 5910005\end{array}$\\
  $\psi_{\sss u}$
     & 1/2 & $({\bf 3}, {\bf 1},  \frac23)$ & no & {\tt YF3u}
     & $\begin{array}{l l l} 5920002 & 5920004 & 5920006\end{array}$\\
  $\psi_{\sss d}$
     & 1/2 & $({\bf 3}, {\bf 1}, -\frac13)$ & no &{\tt YF3d}
     & $\begin{array}{l l l} 5920001 & 5920003 & 5920005\end{array}$\\
\end{tabular}
\caption{\it New particles supplementing the Standard Model field content, given
 together with their representations under $SU(3)_c\times SU(2)_L\times U(1)_Y$,
 their Majorana nature,
 their name in the \fr\ implementation and the associated Particle Data Group
 (PDG) identifiers. Three generations of mediators (second part of the table)
 are included.}
\label{tab:fld}
\end{table*}

We consider a generic $t$-channel dark matter simplified model in which the
Standard Model (SM) is extended by several incarnations of two extra fields, a
dark matter candidate (that we generically denote by $X$) and a mediator lying
in the fundamental representation of $SU(3)_c$ (that we generically denote by
$Y$). In order to maintain the model as general as possible, we allow
for several options for the spin of the new particles and therefore include
six new dark matter fields $\tilde S$, $S$, $\tilde\chi$, $\chi$, $\tilde V_\mu$
and $V_\mu$, all lying in the singlet representation of the SM gauge group
$SU(3)_c\times SU(2)_L\times U(1)_Y$. These fields respectively correspond to a
real scalar field, a complex scalar field, a Majorana spinor, a Dirac spinor, a
real vector field and a complex vector field.

The most general Lagrangian
embedding all the interactions of these fields with the SM can be written, after
imposing that electroweak gauge invariance is preserved, as%
\be\bsp
 \lag = &\ \lag_{\rm SM} + \lag_{\rm kin}
   + \lag_F(\chi) + \lag_F(\tilde\chi)
  \\ &\
   + \lag_S(S)    + \lag_S(\tilde S)
   + \lag_V(V)    + \lag_V(\tilde V) ,
\esp\ee%
where $\lag_{\rm SM}$ is the SM Lagrangian and $\lag_{\rm kin}$ contains
gauge-invariant kinetic and mass terms for all new fields. The
fermionic, scalar and vector dark matter Lagrangi\-ans read%
\be\bsp
  \lag_F(X) = &\ \Big[
          {\bf \lambda_{\sss Q}}   \bar X\qL \varphi^\dag_{\sss Q}
    \!+\! {\bf \lambda_{\sss u}} \bar X\uR \varphi^\dag_{\sss u}
    \!+\! {\bf \lambda_{\sss d}} \bar X\dR \varphi^\dag_{\sss d}
    \!+\! {\rm h.c.} \Big] \ ,\\
  \lag_S(X) = &\ \Big[
         {\bf \hat\lambda_{\sss Q}} \bar\psi_{\sss Q} \qL X
   \!+\! {\bf \hat\lambda_{\sss u}} \bar\psi_{\sss u} \uR X
   \!+\! {\bf \hat\lambda_{\sss d}} \bar\psi_{\sss d} \dR X
   \!+\! {\rm h.c.} \Big] \ , \\
  \lag_V(X) = &\ \Big[
         {\bf \hat\lambda_{\sss Q}} \bar\psi_{\sss Q} \slashed{X} \qL
   \!+\! {\bf \hat\lambda_{\sss u}} \bar\psi_{\sss u} \slashed{X} \uR
   \!+\! {\bf \hat\lambda_{\sss d}} \bar\psi_{\sss d} \slashed{X} \dR
   \!+\! {\rm h.c.} \Big] \ .
\esp \label{eq:lagX}\ee%
In our
notation, $\qL$ stands for the $SU(2)_L$ doublet of left-handed quarks and $\uR$
and $\dR$ are the up-type and down-type $SU(2)_L$ singlets of right-handed
quarks respectively. The scalar mediators $\varphi_{\sss Q}$, $\varphi_{\sss u}$
and $\varphi_{\sss d}$ are chosen to solely interact with the $\qL$, $\uR$ and
$\dR$ quarks, as for the fermionic mediators $\psi_{\sss Q}$, $\psi_{\sss u}$
and $\psi_{\sss d}$ (that are thus vector-like). The mediators therefore lie
in the same SM representation as their quark partners. In the above expression,
we have understood all flavour indices for clarity. The ${\bf
\lambda_{\sss Q}}$, ${\bf \lambda_{\sss u}}$ and ${\bf \lambda_{\sss d}}$
coupling strengths are hence $3\times 3$ matrices in the flavour space, that we
moreover consider real and flavour-diagonal for simplicity.

\begin{table}
\centering
\renewcommand{\arraystretch}{1.4}
\setlength\tabcolsep{10pt}
\begin{tabular}{c c c c}
  Coupling & \fr\ name & LH block\\
  \hline\hline
  $(\lambda_{\sss Q})_{ij}$ & {\tt lamS3Q} & {\tt DMS3Q}\\
  $(\lambda_{\sss u})_{ij}$ & {\tt lamS3u} & {\tt DMS3U}\\
  $(\lambda_{\sss d})_{ij}$ & {\tt lamSdD} & {\tt DMS3D}\\
  $(\hat\lambda_{\sss Q})_{ij}$ & {\tt lamF3Q} & {\tt DMF3Q}\\
  $(\hat\lambda_{\sss u})_{ij}$ & {\tt lamF3u} & {\tt DMF3U}\\
  $(\hat\lambda_{\sss d})_{ij}$ & {\tt lamF3d} & {\tt DMF3D}\\
\end{tabular}
\caption{\it New couplings dictating the interactions of the new particles with
  the Standard Model sector. Each coupling is given together with the
  associated \fr\ symbol and the Les Houches (LH) block of the parameter card.}
\label{tab:prm}
\end{table}

The new physics particles of the simplified model are given in
table~\ref{tab:fld}, together with their representation under the gauge and
Poincar\'e groups, their potential Majorana nature, the adopted particle name in
the \fr\ implementation and the adopted Particle Data Group (PDG)
identifiers~\cite{Tanabashi:2018oca}. The conventions for the different coupling
parameters are summarised in table~\ref{tab:prm}, in which they are given
together with the name used in the \fr\ implementation and the Les Houches (LH)
blocks~\cite{Skands:2003cj} storing their numerical values when running tools
like \mg\ or \maddm.

By relying on a joint usage of the \fr~\cite{Alloul:2013bka}, \nloct~\cite{%
Degrande:2014vpa} and \fa~\cite{Hahn:2000kx} packages, we automatically generate
a UFO model~\cite{Degrande:2011ua} that can be used by \mg\ for both leading
order (LO) and NLO
computations. This UFO model includes all UV counterterms allowing for the
renormalisation of the model with respect to the QCD interactions, as well as
all $R_2$ Feynman rules that are relevant for the numerical evaluation of
one-loop integrals in four dimensions.

\begin{table}
\centering
\renewcommand{\arraystretch}{1.4}
\setlength\tabcolsep{9pt}
\begin{tabular}{c c c c}
  Name & DM & Mediators & Parameters\\
  \hline\hline
  {\tt S3M\_uni} & $\tilde\chi$ &
   \multirow{2}{*}{$\varphi_{\sss Q_f}$, $\varphi_{\sss u_f}$, $\varphi_{\sss d_f}$} &
   \multirow{6}{*}{$M_\varphi$, $M_\chi$, $\lambda_{\sss \varphi}$}\\
  {\tt S3D\_uni} & $\chi$\\
  \cdashline{1-3}
  {\tt S3M\_3rd}& $\tilde\chi$ &
   \multirow{2}{*}{$\varphi_{\sss Q_3}$, $\varphi_{\sss u_3}$, $\varphi_{\sss d_3}$}\\
  {\tt S3D\_3rd} & $\chi$\\
  \cdashline{1-3}
  {\tt S3M\_uR} & $\tilde\chi$ &\multirow{2}{*}{$\varphi_{\sss u_1}$} \\
  {\tt S3D\_uR} & $\chi$\\
  \hline
  {\tt F3S\_uni}& $\tilde S$ &
   \multirow{2}{*}{$\psi_{\sss Q_f}$, $\psi_{\sss u_f}$, $\psi_{\sss d_f}$} &
   \multirow{6}{*}{$M_S$, $M_\psi$, $\hat\lambda_{\sss \psi}$}\\
  {\tt F3C\_uni} & $S$\\
  \cdashline{1-3}
  {\tt F3S\_3rd}& $\tilde S$& \multirow{2}{*}{$\psi_{\sss Q_3}$, $\psi_{\sss u_3}$, $\psi_{\sss d_3}$}\\
  {\tt F3C\_3rd} & $S$\\
  \cdashline{1-3}
  {\tt F3S\_uR}& $\tilde S$& \multirow{2}{*}{$\psi_{\sss u_1}$}\\
  {\tt F3C\_uR} & $S$ \\
  \hline
  {\tt F3V\_uni}& $\tilde V_\mu$&
   \multirow{2}{*}{$\psi_{\sss Q_f}$, $\psi_{\sss u_f}$, $\psi_{\sss d_f}$} &
   \multirow{6}{*}{$M_V$, $M_\psi$, $\hat\lambda_{\sss \psi}$}\\
  {\tt F3W\_uni}& $V_\mu$ \\
  \cdashline{1-3}
  {\tt F3V\_3rd}& $\tilde V_\mu$& \multirow{2}{*}{$\psi_{\sss Q_3}$, $\psi_{\sss u_3}$, $\psi_{\sss d_3}$}\\
  {\tt F3W\_3rd}& $V_\mu$\\
  \cdashline{1-3}
  {\tt F3V\_uR}& $\tilde V_\mu$& \multirow{2}{*}{$\psi_{\sss u_1}$}\\
  {\tt F3W\_uR}& $V_\mu$\\
\end{tabular}
\caption{\it List of all restrictions included in the \dmsimpt\ UFO model. In
  each case, the simplified model contains a single class of
  mass-degenerate mediators (where $f=1,2,3$ is a flavour index), a specific
  dark matter candidate and universal and flavour-conserving dark matter
  couplings $\lambda_{\sss \varphi}$ and $\hat\lambda_{\sss\psi}$.}
\label{tab:restr}
\end{table}

The model is shipped with a large ensemble of restrictions dedicated to
specific $t$-channel simplified models. These are summarised in
table~\ref{tab:restr} whe\-re for each restriction, we specify the active new
physics states, all other states being taken decoupled and non-interacting. In
other words, each restriction consists in a simplified model in which the SM is
extended by a specific class of mediators, and a given dark matter
state. In order to reduce the number of free parameters, all (active) mediators
are taken mass-degenerate. A given restriction named \verb+XYZ+ can be
loaded in \mg\ (or \maddm) by typing, within the tool command line interface,
\begin{verbatim} import model DMSimp_t-XYZ --modelname \end{verbatim}

In the model restrictions whose name ends with the {\tt uni} suffix, all
twelve flavours of mediators are considered, their mass and interaction
strengths being taken flavour-conserving and universal,%
\be
  ({\bf \lambda_{\sss F}})_{ij} = \lambda_{\sss \varphi} \delta_{ij}
  \qquad\text{and}\qquad
  ({\bf \hat \lambda_{\sss F}})_{ij} = \hat\lambda_{\sss \psi} \delta_{ij} \ ,
\ee%
for $F=Q$, $u$ and $d$. In model restrictions of the {\tt uR} class, only
mediators coupling to the right-handed up quark are taken as active,%
\be
  ({\bf \lambda_{\sss u}})_{11} = \lambda_{\sss \varphi}
  \qquad\text{and}\qquad
  ({\bf \hat \lambda_{\sss u}})_{11} = \hat\lambda_{\sss \psi} \ ,
\ee%
all other couplings being vanishing, whilst in the {\tt 3rd} class of model
restrictions, we only consider the mediator coupling to the third
generation of SM quarks,%
\be\bsp
  & ({\bf \lambda_{\sss Q}})_{33} =
    ({\bf \lambda_{\sss u}})_{33} =
    ({\bf \lambda_{\sss d}})_{33} = \lambda_{\sss \varphi} \ , \\
  & ({\bf \hat \lambda_{\sss Q}})_{11} =
    ({\bf \hat \lambda_{\sss u}})_{22} =
    ({\bf \hat \lambda_{\sss d}})_{33} = \hat\lambda_{\sss \psi} \ ,
\esp \ee%
all other couplings being again assumed vanishing.

\subsection{The {\tt S3M}/{\tt S3D} class of models}\label{sec:s3ms3d}
In {\tt S3M}-type and {\tt S3D}-type models, the dark matter is taken to
be respectively the Majorana and Dirac state $\tilde\chi$ and $\chi$ of mass
$M_\chi$. As written in section~\ref{sec:general}, all mediators are considered
degenerate of mass $M_\varphi$, and all new
physics interactions are universal and flavour-conserving with a global strength
$\lambda_{\sss\varphi}$. The generic Lagrangian $\lag_F$ of
eq.~\eqref{eq:lagX} therefore simplifies to%
\be
 \lag_{\tt X\_uni}(X) = \sum_{F=Q,u,d}~\sum_{f=1}^3
  \Big[\lambda_{\sss\varphi} \bar X F_f \varphi^\dag_{\sss F_f}
   + {\rm h.c.}\Big] \ ,
\ee%
where $X=\chi$ ({\tt S3D}) or $\tilde\chi$ ({\tt S3M})
equivalently refers to Dirac or Majorana dark matter, and $f$ is a
generation index. The model is thus defined by three parameters,%
\be
  \big\{ M_\chi, \ M_\varphi, \ \lambda_{\sss\varphi} \big\} \ .
\label{eq:S3Mprm}\ee%
In the universal {\tt S3M\_uni} and {\tt S3D\_uni} restrictions, the simplified
model includes all twelve mediators, whilst in the {\tt S3M\_3rd} and
{\tt S3D\_3rd} restrictions, the setup is further simplified and dark matter
only couples to the third generation via the four corresponding mediators.
In the {\tt S3M\_uR} and {\tt S3D\_uR} restrictions, only a coupling to the
right-handed up quark $u_1$ is considered, through a single mediator.
The associated Lagrangians read,%
\be\bsp
 \lag_{\tt X\_3rd}(X)= &\ \sum_{F=Q,u,d}\Big[
   \lambda_{\sss\varphi} \bar X F_3 \varphi^\dag_{\sss F_3}
   + {\rm h.c.} \Big] \ ,\\
   \lag_{\tt X\_uR}(X) = &\
    \Big[\lambda_{\sss\varphi} \bar X\uR_1\varphi^\dag_{\sss u_1}
   + {\rm h.c.} \Big] \ .
\esp\ee%

\subsection{The {\tt F3S}/{\tt F3C} class of models}\label{sec:f3s}
In {\tt F3S}-type and {\tt F3C}-type models, the dark matter consists of the
real and complex scalar state $\tilde S$ and $S$ of mass $M_S$ respectively. As
in the previous subsection, all mediators are assumed to
be degenerate of mass $M_\psi$, and all new
physics interactions are universal and flavour-conserving with a strength
$\hat\lambda_{\sss \psi}$. The Lagrangian $\lag_S$ of
eq.~\eqref{eq:lagX} therefore simplifies to%
\be
 \lag_{\tt X\_uni}(X) =
     \sum_{F=Q,u,d}~\sum_{f=1}^3\Big[
     {\bf \hat\lambda_{\sss \psi}} \bar\psi_{\sss F_f} F_f X
   + {\rm h.c.} \Big] \ ,
\ee%
where $X=\tilde S$ ({\tt F3S}) and $S$ ({\tt F3C}) in the real and complex case.
The model is defined by three parameters,%
\be
  \big\{ M_S, \ M_\psi, \ \hat\lambda_{\sss\psi} \big\} \ .
\ee%
In the universal {\tt F3S\_uni} and {\tt F3C\_uni}
restrictions, dark matter couples to all SM quark eigenstates through twelve
mediators. In the third generation {\tt F3S\_3rd} and {\tt F3C\_3rd} models, its
couplings are restricted to the bottom and top quark ones and the corresponding
four mediators, while in the {\tt F3S\_uR} and {\tt F3C\_uR} models, dark matter
only couples to the right-handed up quark. The associated Lagrangians are%
\be\bsp
 \lag_{\tt X\_3rd}(X) = &\ \sum_{F=Q,u,d}\Big[
     {\bf \hat\lambda_{\sss \psi}} \bar\psi_{\sss F_3} F_3 X
   + {\rm h.c.} \Big] \ , \\
 \lag_{\tt X\_uR}(X) = &\ \Big[
     {\bf \hat\lambda_{\sss \psi}} \bar\psi_{\sss u_1} \uR_1 X
   + {\rm h.c.} \Big] \ .
\esp\ee%

\subsection{The {\tt F3V}/{\tt F3W} class of models}\label{sec:f3v}
In the {\tt F3V} and {\tt F3W} types of models, the dark matter is a real and
complex vector state $\tilde V_\mu$ and $V_\mu$ of mass $M_V$ respectively. All
mediators are
degenerate of mass $M_\psi$, and all new physics interactions are universal and
flavour-conserving with a common strength $\hat\lambda_{\sss \psi}$. The
Lagrangian $\lag_V$ of eq.~\eqref{eq:lagX} is simplified to%
\be
 \lag_{\tt X\_uni}(X) = \sum_{F=Q,u,d}~\sum_{f=1}^3\Big[
     {\bf \hat\lambda_{\sss \psi}} \bar\psi_{\sss F_f} \slashed{X} F_f
   + {\rm h.c.} \Big] \ ,
\ee%
where $X = \tilde V$ ({\tt F3V}) or $V$ ({\tt F3W}) in the real and complex
vector case. The model is defined by three parameters,%
\be
  \big\{ M_V, \ M_\psi, \ \hat\lambda_{\sss\psi} \big\} \ .
\ee%
In the {\tt F3V\_uni} and {\tt F3W\_uni} restrictions, all twelve mediators are
included. In contrast, in the {\tt F3V\_3rd} and {\tt F3W\_3rd} restrictions,
only the four mediators relating dark matter to the top and bottom quarks are
included, whilst in the {\tt F3V\_uR} and {\tt F3W\_uR} models, the only
non-vanishing coupling is the one to the right-handed up-quark. The associated
Lagrangians read%
\be\bsp
 \lag_{\tt X\_3rd}(X) = &\ \sum_{F=Q,u,d}\Big[
     {\bf \hat\lambda_{\sss \psi}} \bar\psi_{\sss F_3}\slashed{X} F_3
   + {\rm h.c.} \Big] \ , \\
 \lag_{\tt F3S\_uR}(X) = &\ \Big[
     {\bf \hat\lambda_{\sss \psi}} \bar\psi_{\sss u_1}\slashed{X}\uR_1
   + {\rm h.c.} \Big] \ .
\esp\ee%

\section{Matching NLO QCD fixed-order calculations with parton showers}
\label{sec:nlomatch}

\subsection{Generalities}
In the class of simplified models under consideration, the computation of
NLO QCD corrections involve real emission diagrams possibly featuring
intermediate $s$-channel resonances. These should be treated consistently in
order not to apparently spoil the convergence of the perturbative series by
yielding NLO cross sections much larger than the associated LO ones. This
occurs when the cross section related to the production of the resonant state
is much larger than the one of the initially considered process. Moreover, we
aim at combining events describing all possible new physics
processes of a given model at the NLO accuracy in QCD. We will hence
consider the production of a pair of dark matter particles ($pp\to XX$), of any
mediators ($pp\to Y_iY_j$), as well as the associated production of a mediator
and a dark matter state ($pp\to XY_i$). Therefore, the subtraction of all
resonant contributions in the real corrections is mandatory to avoid their
double-counting when combining the three types of processes.

Different strategies dealing with the treatment of these resonances have been
recently automated within the \mg\ framework~\cite{Frixione:2019fxg}. They
include diagram removal methods with or without keeping the interferences
between the resonant and non-resonant contributions~\cite{Hollik:2012rc}, as
well as various techniques to subtract the resonant
contribution from the full amplitude~\cite{Frixione:2008yi}. In the
following, we employ one of such strategies, in which all squared resonant
diagram contributions are discarded whilst the interferences of the resonant and
non-resonant diagrams are kept. All available methods should however lead
to numerically similar results if the resonant process can be consistently
defined.

In practice, \mg\ has to be run together with
the \mstr\ plugin\footnote{The \mstr\ plugin can be downloaded from
\url{https://code.launchpad.net/~maddevelopers/mg5amcnlo/MadSTRPlugin}} that can
be activated by starting \mg\ as%
\begin{verbatim}
 mg5_aMC --mode=MadSTR
\end{verbatim}%
The code is then used to simulate events, at the NLO accuracy in QCD, relevant
for all new physics processes allowed by $t$-channel dark matter models. The
considered processes can be classified into three categories,%
\be\bsp
  p p &\to XX\ , \\
  p p & \to XY\ \text{with}\  Y \to X j \ ,\\%\qquad\text{and}\qquad
  p p & \to YY\ \text{with}\  Y \to X j \ .
\esp\label{eq:processes}
\ee
This corresponds to the production of a pair of dark matter particles
(generically denoted by $XX$), the associated production of a mediator and a
dark matter particle (generically denoted by $XY$), and the production of a pair
of mediators (generically denoted by $YY$). In the latter two
cases, the mediator further decays into a SM quark and dark matter.

After simulating each process separately, the different contributions are
combined, which is only possible if all resonant pieces from the real emission
to the three subprocesses are subtracted. For instance, the diagrams associated
with the second Born subprocess ($pp \to XY \to XXj$) are included in those
related to the real corrections to the first subprocess ($pp \to XX$). In order
to avoid any double counting, we include the resonant component into the Born
contribution to $XY$ production, and the non-resonant one into the real
corrections to $XX$ production.

We import the \dmsimpt\ UFO model in \mg\ to deal with the generation of
hard-scattering events at the NLO accuracy for all the processes of
eq.~\eqref{eq:processes}, using the \mstr\ plugin and convoluting the matrix
elements with the NLO set of NNPDF 3.0 parton distribution functions
(PDFs)~\cite{Ball:2014uwa}
accessed via the LHAPDF~6 library~\cite{Buckley:2014ana}. Mediator decays are
handled with the \msp~\cite{Artoisenet:2012st} and \mwdth~\cite{Alwall:2014bza}
program\-mes, which allows for the factorisation of the production and decay
processes in a way retaining both off-shell and spin correlation effects.
The resulting partonic events are matched with parton showers as described by
the \py\ package~\cite{Sjostrand:2014zea}, following the MC@NLO
prescription~\cite{Frixione:2002ik}. We also use \py\ to handle
hadronisation. We then reconstruct the hadron-level events by clustering hadrons
according to the anti-$k_T$ algorithm with a separation parameter set to
$\Delta R=0.4$~\cite{Cacciari:2008gp}, as implemented in the \fj\
software~\cite{Cacciari:2011ma} that we drive from \ma~\cite{Conte:2012fm,%
Conte:2018vmg}. The latter programme is also used for the
generation of the differential distributions studied in
section~\ref{sec:mctruth}, and the reinterpretation analysis of the LHC results
in section~\ref{sec:recast}.

\subsection{Simulating an {\tt S3D\_uR} dark matter signal}
\label{sec:simu}
\begin{figure}
  \centering
  \includegraphics[trim={0 0 50 0} ,clip,width=.35\columnwidth]{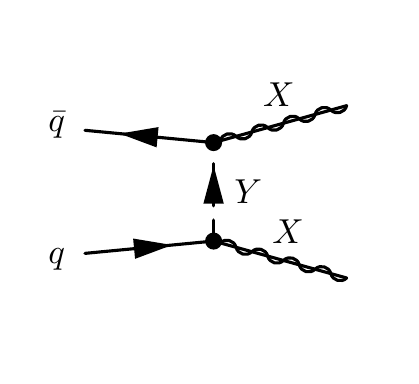}
  \includegraphics[trim={0 0  0 30},clip,width=.48\columnwidth]{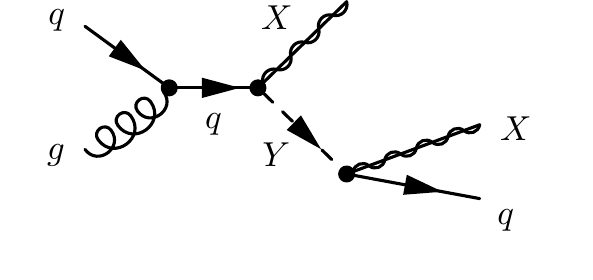}
  \caption{
    \label{fig:xx_diags}
    \it Representative LO Feynman diagrams describing the
    production of a pair of dark matter particles (left) and the associated
    production of a mediator with a dark matter state (right). The mediator
    decay into dark matter and a quark is included.}
  \includegraphics[width=.48\columnwidth]{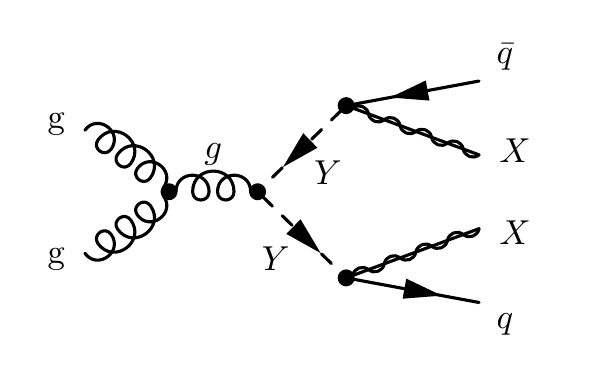}
  \includegraphics[width=.48\columnwidth]{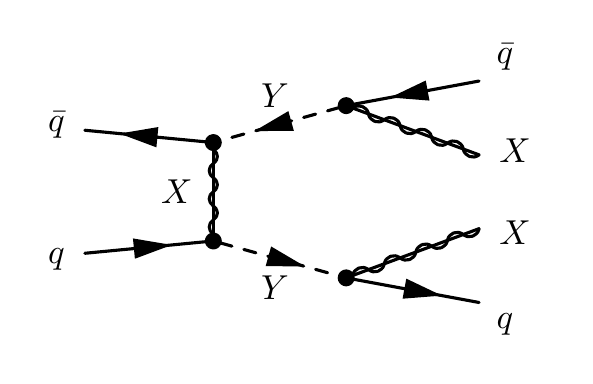}
  \caption{
    \label{fig:yy_diag}
    \it Same as figure~\ref{fig:xx_diags} but for mediator pair-production (and
    decay) in the QCD (left) and $t$-channel dark matter exchange (right)
    channels.}
\end{figure}

In the following, we illustrate how NLO predictions matched with parton showers
can be achieved in the {\tt S3D\_uR} class of model. In the $X/Y$ notations of
eq.~\eqref{eq:processes}, we thus have, $X=\chi,\bar\chi$ and
$Y=\varphi_{\sss u_1},\varphi_{\sss u_1}^\dag$.

Events originating from dark
matter pair production at NLO ($p p \to XX$; see \eg\ the left
panel of figure~\ref{fig:xx_diags} for a representative LO Feynman diagram) are
generated by
starting \mg\ with the \mstr\ plugin switched on. The usual {\tt generate} and
{\tt output} commands available from the \mg\ command line interface~\cite{%
Alwall:2014hca} are then cast, after having imported the restricted UFO model,%
\begin{verbatim}
 import model DMSimpt-S3D_uR --modelname
 generate p p > xd xd~ / yf3qu1 yf3qu2      \
  yf3qu3 yf3qd1 yf3qd2 yf3qd3 yf3u1  yf3u2  \
  yf3u3  yf3d1  yf3d2  yf3d3  ys3qu1 ys3qu2 \
  ys3qu3 ys3qd1 ys3qd2 ys3qd3 ys3u2  ys3u3  \
  ys3d1 ys3d2 ys3d3 xs xm xv [QCD]
 output
\end{verbatim}%

In order for the restricted model to be dealt with consistently, it is crucial
to explicitly forbid any decoupled particle to run into any virtual loop. This
is implemented at the level of the {\tt generate} command, in which we manually
exclude all fermionic mediators, all scalar mediators not coupling to the
right-handed up quark $u_R$ and all irrelevant dark matter states of
the model. The UFO conventions for the particle names follow the \fr\ ones
introduced in table~\ref{tab:fld} (we recall that the \mg\ command line
interface is case insensitive), additionally including an integer number
for the generation indices. On run time, the \mstr\ plugin takes care of
identifying and
treating any potentially resonant contribution. In our case, the squared
resonant contributions are discarded, whilst the interferences of the resonance
with the non-resonant continuum are kept.

Events describing the associated production of a mediator with a dark matter
particle ($pp\to XY$; see \eg\ the right panel of figure~\ref{fig:xx_diags} for
a representative diagram including the mediator decay process) are generated in
a similar fashion,%
\begin{verbatim}
 import model DMSimpt-S3D_uR --modelname
 define dm = xd xd~
 define yy1 = ys3u1 ys3u1~
 generate p p > dm yy1 / yf3qu1 yf3qu2      \
  yf3qu3 yf3qd1 yf3qd2 yf3qd3 yf3u1  yf3u2  \
  yf3u3  yf3d1  yf3d2  yf3d3  ys3qu1 ys3qu2 \
  ys3qu3 ys3qd1 ys3qd2 ys3qd3 ys3u2  ys3u3  \
  ys3d1 ys3d2 ys3d3 xs xm xv [QCD]
 output
\end{verbatim}%
where we make use of the {\tt dm} and {\tt yy1} multiparticle labels to
guarantee that all potential particle/antiparticle combinations are accounted
for.

Mediator pair production is generally dominated by QCD contributions that are
independent of the dark matter mass and couplings, as illustrated by the first
Feynman diagram of figure~\ref{fig:yy_diag} that also includes the mediator
decay process. However, if $\lambda_{\sss\varphi}$ is large enough,
$t$-channel dark matter exchanges, as depicted by the second diagram of
figure~\ref{fig:yy_diag}, could significantly contribute. NLO QCD corrections to
the strong contributions to mediator pair-production of ${\cal O}(\as)$ can be
straightforwardly calculated by typing in the \mg\ command line interface,%
\begin{verbatim}
 import model DMSimpt-S3D_uR --modelname
 define yy1 = ys3u1 ys3u1~
 generate p p > yy1 yy1 / yf3qu1 yf3qu2     \
  yf3qu3 yf3qd1 yf3qd2 yf3qd3 yf3u1  yf3u2  \
  yf3u3  yf3d1  yf3d2  yf3d3  ys3qu1 ys3qu2 \
  ys3qu3 ys3qd1 ys3qd2 ys3qd3 ys3u2  ys3u3  \
  ys3d1 ys3d2 ys3d3 xs xm xv [QCD]
 output
\end{verbatim}%
without using the \mstr\ plugin. Making use of the coupling order
information of the model, \mg\ automatically restricts the process to its pure
QCD contribution, neglecting any $t$-channel dark matter exchange at the Born
level and any real emission or virtual contribution depending on
$\lambda_{\sss\varphi}$. The considered Born contribution is thus of
${\cal O}(\as^2)$ whilst the NLO component is of ${\cal O}(\as^3)$ and free
of any resonance.

The pure $t$-channel contribution can be evaluated, at the NLO accuracy in QCD,
by typing%
\begin{verbatim}
 import model DMSimpt-S3D_uR --modelname
 define yy1 = ys3u1 ys3u1~
 generate p p > yy1 yy1 DMT=2 QCD=0 QED=0   \
  yf3qu1 yf3qu2 yf3qu3 yf3qd1 yf3qd2 yf3qd3 \
  yf3u1  yf3u2  yf3u3  yf3d1  yf3d2  yf3d3  \
  ys3qu1 ys3qu2 ys3qu3 ys3qd1 ys3qd2 ys3qd3 \
  ys3u2  ys3u3  ys3d1 ys3d2 ys3d3 xs xm xv  \
  [QCD]
 output
\end{verbatim}%
The coupling order restriction {\tt DMT=2 QCD=0 QED=0} attached to the
{\tt generate} command guarantees that the Born amplitude is proportional to
$\lambda_{\sss \varphi}^2$ and does not include any contribution depending on
$\as$ or on the electroweak coupling $\alpha$. In other words, any tree-level
diagram including a gluon, a photon or a $Z$-boson propagator is discarded, so
that the Born matrix element is of ${\cal O}(\lambda_{\sss\varphi}^4)$ and the
NLO corrections of ${\cal O}(\lambda_{\sss\varphi}^4 \as)$. In order
to deal with the resonant contributions potentially arising at NLO, the \mstr\
plugin is used.

Care must be taken when dealing with the mixed-order interferences of the QCD
diagrams with the $t$-channel ones. The version 3.x.y of \mg\ being incompatible
with \mstr\ and the version 2.6.x of the code being unable to handle mixed
orders, there is not any publicly available and user-friendly option.
One possible way to cure this issue would be to include in
the UFO model all UV counterterms and $R_2$ rational terms necessary for
mixed-order NLO calculations in QCD, QED and in the new physics $\lambda$
coupling, and to implement in \mfks~\cite{Frederix:2009yq}
all necessary subtraction terms. This
however goes beyond the scope of this work.

We therefore adopt the strategy of
simulating the interferences at LO, and reweight the events
by a $K$-factor assumed to approximate the effect of the QCD corrections.
We multiply the interference event weights by the geometric mean of the
pure QCD and pure $t$-channel $K$-factors, those two NLO to LO ratios being
defined differentially. In other words, each distribution will be reweighted bin
by bin. Event simulation for the interferences is then performed by typing, in
the \mg\ command line interface,%
\begin{verbatim}
 import model DMSimpt-S3D_uR --modelname
 define yy1 = ys3u1 ys3u1~
 generate p p > yy1 yy1 DMT^2==2 / yf3qu1    \
  yf3qu2 yf3qu3 yf3qd1 yf3qd2 yf3qd3 yf3u1  \
  yf3u2  yf3u3  yf3d1  yf3d2  yf3d3  ys3qu1 \
  ys3qu2 ys3qu3 ys3qd1 ys3qd2 ys3qd3 ys3u2  \
  ys3u3  ys3d1 ys3d2 ys3d3 xs xm xv
 output
\end{verbatim}%

\subsection{Total and differential cross sections for {\tt S3D\_uR} dark matter}
\label{sec:mctruth}

\begin{table*}
\centering
\renewcommand{\arraystretch}{1.7}
\setlength\tabcolsep{8pt}
\begin{tabular}{c | c | c | c c c}
  Scen. & $XX$ [fb] & $XY$ [fb] & $YY$ (total) [fb] &  $YY$ (QCD) [fb] &
   $YY$ ($t$-channel) [fb]\\
  \hline\hline
  {\bf S1} &  $775.3_{-0.8\%}^{+0.4\%}\pm1.9\%$ &
    $1617_{-13.4\%}^{+16.5\%}\pm1.0\%$ &
    $473.5_{-16.9\%}^{+23.6\%}\pm3.0\%$&
    $324.2_{-23.8\%}^{+34.2\%}\pm3.4\%$&
    $261.5_{-6.3\%}^{+7.1\%}\pm2.5\%$\\
  {\bf S2} & $122.0_{-2.0\%}^{+1.8\%}\pm1.9\%$ &
    $74.1_{-15.8\%}^{+20.3\%}\pm1.2\%$ &
    $7.452_{-14.5\%}^{+19.8\%}\pm5.6\%$&
    $3.545_{-25.4\%}^{+37.3\%}\pm7.2\%$&
    $6.939_{-9.4\%}^{+11.1\%}\pm5.0\%$\\
  \hline
  {\bf S1} & $929.8_{-1.3\%}^{+1.9\%}\pm1.9\%$ &
    $2212_{-6.3\%}^{+5.9\%}\pm1.0\%$ &
    $648.4_{-9.2\%}^{+8.0\%}\pm3.1\%$ &
    $484.7_{-12.4\%}^{+10.7\%}\pm3.4\%$ &
    $314.1_{-2.6\%}^{+2.6\%}\pm2.5\%$\\
  {\bf S2} & $139.1_{-1.1\%}^{+1.3\%}\pm2.0\%$ &
    $101.8_{-7.1\%}^{+6.0\%}\pm1.2\%$ &
    $9.888_{-7.6\%}^{+6.5\%}\pm5.8\%$ &
    $5.303_{-13.3\%}^{+11.2\%}\pm7.4\%$ &
    $8.749_{-3.9\%}^{+3.6\%}\pm4.9\%$
\end{tabular}
\caption{\it Total cross sections at LO (upper) and NLO (lower), in fb, for the
  subprocesses of eq.~\eqref{eq:processes} and the benchmark scenarios defined
  in eqs.~\eqref{eq:benchmarks} and \eqref{eq:benchmarks_2}. Our predictions are
  given together with the scale and parton density uncertainties.}
\label{tab:xsec}
\end{table*}

In order to illustrate how all subprocesses of eq.~\eqref{eq:processes} could
impact a dark matter signal at the LHC, we consider two benchmark scenarios
representative of the {\tt S3D\_uR} model. We fix the dark matter and mediator
masses to%
\be\bsp
  {\bf S1.} \quad&M_\chi=150~{\rm GeV}\ , \qquad M_\varphi = 500~{\rm GeV} \ ,\\
  {\bf S2.} \quad&M_\chi=150~{\rm GeV}\ , \qquad M_\varphi = 1000~{\rm GeV} \ ,
\esp\label{eq:benchmarks}\ee%
and the new physics coupling to%
\be
\lambda_{\sss \varphi} = 1 \ .
\label{eq:benchmarks_2}\ee%
In the first scenario {\bf S1}, the spectrum is more compressed although there
is enough phase space for the light mediator to decay into a dark matter
particle and a hard jet. In the second scenario {\bf S2}, the mediator is
heavier, its mass being fixed to a more realistic value with respect to current
squark mass limits~\cite{Sirunyan:2019xwh,ATLAS-CONF-2019-040}. Whilst present
supersymmetry bounds on the strongly-interacting superpartners are usually
stricter, they are not directly applicable to our setup by virtue of the
different nature of the dark matter and mediator particles. We therefore ignore
them for now and address this point in section~\ref{sec:recast}.

In table~\ref{tab:xsec}, we present total cross sections for the various
processes of eq.~\eqref{eq:processes} and for the two considered benchmarks,
both at LO (upper panel) and NLO (lower panel) accuracy, and for a setup in
which the $pp\to XX$ process simulation includes a transverse momentum
($p_T$) cut of 100~GeV on the leading jet at the matrix-element-generator
level. For each of the
subprocesses, the NLO $K$-factor defined as the ratio of the NLO predictions to
the LO one is large. This emphasises the relevance of using rates that are
NLO-accurate to avoid underestimating signal yields. Our LO and NLO
predictions also include theoretical scale uncertainties originating from
missing higher-order corrections and those associated with the parton density
fit. We estimate the former by a nine-point independent scale variation in
which the renormalisation and factorisation scales are varied by a factor of 2
up and down with respect to a central scale set to the average transverse mass
of the final-state particles.

Except for dark-matter pair-production ($pp\to
XX$) that is insensitive to $\alpha_s$ at the lowest order (see \eg\ the left
panel of figure~\ref{fig:xx_diags}), LO
predictions are affected by large scale uncertainties that are significantly
reduced when NLO corrections are included. This consists in the second major
benefit of higher-order calculations: the reduction of the
theoretical systematics. The second source of theoretical uncertainties,
the PDF errors, yields a similar effect at LO and NLO as the same parton density
set has been used. Those errors are reasonably small as our benchmark scenarios
feature masses leading to a moderate Bjorken-$x$ regime. Additionally, we have
verified that the QCD contribution to mediator pair production agrees with the
expectation for squark pair-production in a supersymmetric simplified scenario
in which all superpartners but a single squark are
decoupled~\cite{Frixione:2019fxg}.

In the right panel of the table, we investigate the impact of the QCD and
$t$-channel contributions to the production of a pair of mediators. The adopted
scenarios, with their large coupling choice of eq.~\eqref{eq:benchmarks_2},
ensure that the $t$-channel contribution is relevant and cannot be neglected. On
the contrary, for slightly smaller coupling choices, only QCD production would
remain, as the $t$-channel amplitude squared is
proportional to $\lambda_{\sss\varphi}^4$ and the interference between the QCD
and $t$-channel mode to $\lambda_{\sss\varphi}^2$. For $\lambda_{\sss\varphi} =
1$, the relative
importance of the QCD and $t$-channel modes turns out to differ for the two
benchmark points under consideration. In the case of the {\bf S1} setup, QCD
contributions dominate, as expected for such a small mediator mass $M_\varphi
= 500$~GeV. In contrast, for scenarios like the {\bf S2}
scenario in which the mediator is much heavier, the QCD production mode is
suppressed by virtue of the steeply falling production rate with $M_\varphi$, so
that the $t$-channel contribution dominates. For both cases, the two
contributions are however of a similar order of magnitude and their destructive
interferences are large.

\begin{figure*}
  \centering
  \includegraphics[width=.32\textwidth]{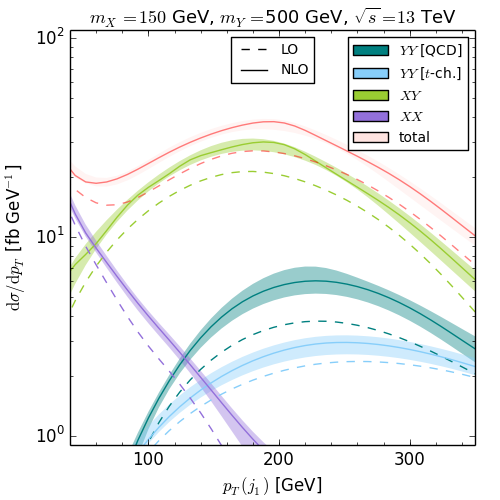}
  \includegraphics[width=.32\textwidth]{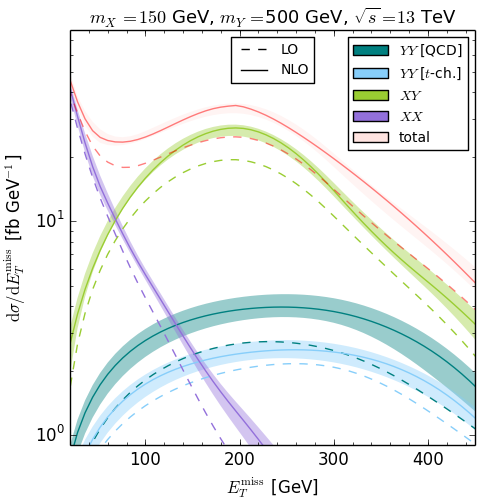}
  \includegraphics[width=.32\textwidth]{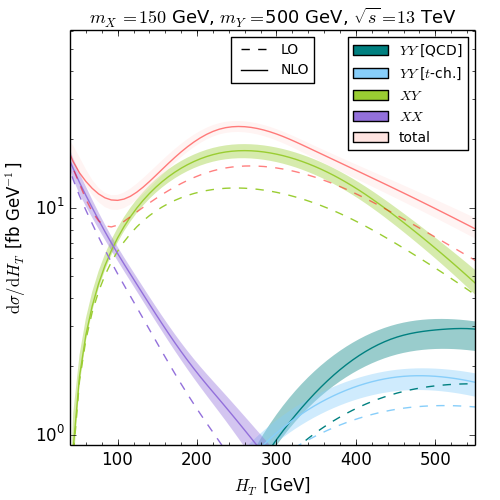}
  \caption{\it
    Selected properties of the new physics signal emerging from the {\bf S1}
    scenario. We present the $p_T$ spectrum of the leading jet (left), as well
    as the $E_T^{\rm miss}$ (central) and $H_T$ (right) distributions. We
    consider the separate contributions of the production of a pair of dark
    matter particles ($XX$; purple), the associated production of a dark matter
    particle and a mediator ($XY$; green), the QCD-induced production of a pair
    of mediators ($YY$ [QCD]; teal) and the $t$-channel-induced production of a
    pair of mediators ($YY$ [$t$-ch.]; blue). The sum of all contributions (red)
    additionally includes the interferences between the two mediator
    pair-production modes. For all channels, we compare NLO predictions (solid
    lines) with LO predictions (dashed lines), and represent the NLO scale
    uncertainty variation bands by shaded areas.
  \label{fig:S1_prop}}\vspace*{.1cm}
  \includegraphics[width=.32\textwidth]{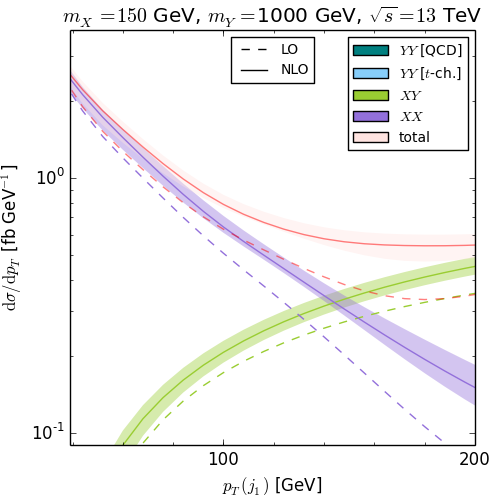}
  \includegraphics[width=.32\textwidth]{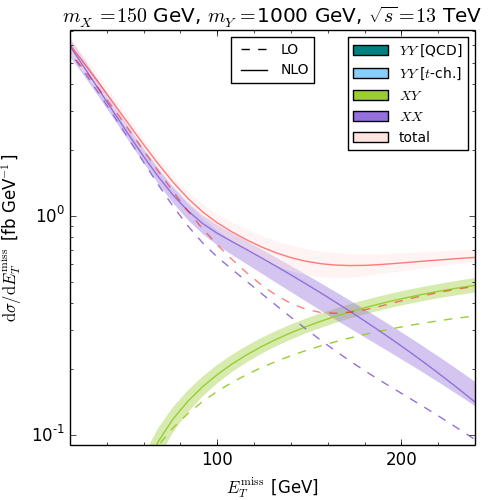}
  \includegraphics[width=.32\textwidth]{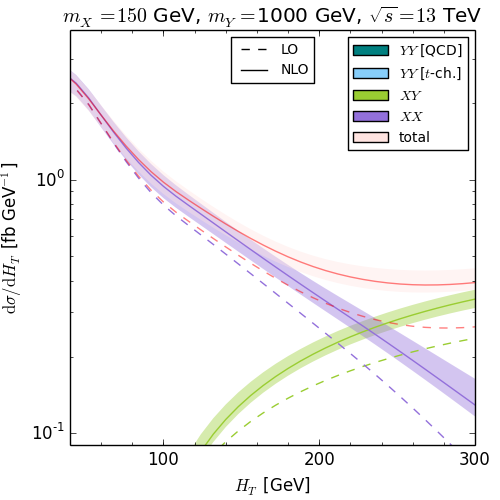}
  \caption{\it Same as figure~\ref{fig:S1_prop} but for the {\bf S2} scenario.
  \label{fig:S2_prop}}
\end{figure*}

The two channels are sensitive to different initial partonic luminosities. QCD
production is mostly induced by gluon fusion (at 80\% and 60\% in the
{\bf S1} and {\bf S2} cases respectively) and $t$-channel production by
quark-antiquark scattering). They feature different PDF uncertainties, as well
as a different dependence on the scales.

In figures~\ref{fig:S1_prop} and \ref{fig:S2_prop}, we present a few properties
of the new physics signal induced by the benchmark scenarios {\bf S1} and
{\bf S2} respectively.
We focus on observables that are relevant for dark matter
searches at the LHC in the monojet channel and consider the description of
the missing transverse energy and jet activity. We show differential
distributions for the transverse momentum of the leading jet $p_T(j_1)$ (left
panel), the missing transverse energy $E_T^{\rm miss}$ (central panel) and the
total hadronic activity $H_T$ (right panel) defined as the scalar sum of the
$p_T$ of all reconstructed jets. For each observable, we present predictions at
LO (dashed
lines) and NLO (solid lines) for the individual contributions of the processes
shown in eq.~\eqref{eq:processes}, as well as for their sum (red). We hence
distinguish the QCD-induced (teal) and $t$-channel-medi\-a\-ted components
(blue) of the mediator pair-production channel ($YY$), the dark-matter pair
production mode ($XX$, purple) and the associated production of a mediator with
dark matter ($XY$, green). The shaded areas around the NLO results
correspond to the uncertainty bands obtained as described above, \ie\ from a
nine-point variation of the unphysical factorisation and renormalisation scales.

Our results show that the dark matter pair-produc\-tion channel, despite a large
production cross section (see table~\ref{tab:xsec}), mainly yields events
featuring a small amount of missing energy and not so much ha\-dro\-nic
activity, even if at the matrix-element level, our simulation includes
a selection cut of 100~GeV on the $p_T$ of the leading
(parton-level) jet. After matching the fixed-order NLO predictions with parton
showers, the emissions originating from this hard parton are often not
reclustered back so that a small ensemble of softer jets are finally
reconstructed from the initial hard radiation. Consequently, the dark matter
particles turn out to be mostly produced back-to-back, which leads to a small
amount of missing energy, and in association with a small hadronic activity. As
a consequence, the efficiency of a typical monojet selection is expected to be
quite reduced as one generally requires a substantial amount of missing energy
and hadronic activity. We refer to section~\ref{sec:recast} for more details.

The leading relevant process from the
cross sections presented in table~\ref{tab:xsec} therefore consists in the
associated production of a dark matter particle with a heavier mediator $pp\to
XY$. The $Y\to X q$ decay of the mediator leads to the production of a
second dark matter state together with a hard parton, which guarantees a much
larger missing transverse energy and hadronic activity than for the $pp\to XX$
channel. The results, depicted by the green curves in figures~\ref{fig:S1_prop}
and \ref{fig:S2_prop} for the {\bf S1} and {\bf S2} scenarios respectively,
confirm this, the corresponding $E_T^{\rm miss}$, $p_T(j_1)$ and
$H_T$ distributions being depleted in the low-energy regime. In the high-energy
tails, the $XY$ contributions moreover almost match entirely the total new
physics signal (red curves), the $pp\to YY$ contributions being only expected to
take over in the very hard part of the phase space (not represented in
figure~\ref{fig:S1_prop} and that is drastically phase-space-suppressed, and
thus not shown, for the {\bf S2} setup in figure~\ref{fig:S2_prop}).
The $XY$ distributions are indeed steeply falling with the energy
scale, compared with the $YY$ ones, so that the $XY$ component of the
signal is only dominant for moderate observable values of a few hundreds
of GeV. The relative importance of the $XY$ process can however be tamed down
by reducing the magnitude of the $\lambda_{\sss\varphi}$ coupling, on which the
normalisation of the distributions depends quadratically as the amplitude of the
corresponding partonic process $qg\to XY$ is linear in both the new physics and
strong coupling constants (see the right diagram in figure~\ref{fig:xx_diags}).

Finally, mediator pair production ($pp\to YY$)
only dominates, as said above, in the harder part of the spectra
where all other contributions are kinematically suppressed. In that regime, the
decays of the two heavy mediators into $Xq$ systems guarantee an amount of
missing energy and hadronic activity greater than for $XY$ production, despite
the global rates being reduced by the large mediator mass. In our analysis, we
distinguish the QCD production mode whose matrix element is
proportional to $\as^2$ and independent of $\lambda_{\sss\varphi}$, and the
$t$-channel one that depends on $\lambda_{\sss\varphi}^4$. Whereas for the
adopted $\lambda_{\sss\varphi}=1$ benchmark value, the two production channels
contribute equivalently, the $t$-channel impact can be
reduced by fixing $\lambda_{\sss\varphi}$ to a smaller value. For
$\lambda_{\sss\varphi}\sim\gw$ ($\gw$ being the weak coupling constant), we
obtain the widely studied supersymmetric limiting scenario in which all
superpartners except the right-handed up squark and a bino-like
neutralino are decoupled, with a difference on the dark matter
nature that is here a Dirac fermion. In this case, only the QCD production of
two mediators matters, and the $t$-channel contribution to $YY$ production and
$XY$ production can be ignored (at least for the considered mediator masses).

\subsection{Collider constraints on {\tt S3D\_uR} dark matter}\label{sec:recast}

\begin{table*}
 \centering
 \renewcommand{\arraystretch}{1.8}
 \setlength\tabcolsep{9pt}
   \begin{tabular}{c c | c c | c c}
     Sc. & Process & CL$_s$ [LO] & $E_T^{\rm miss}$ constraint&
       CL$_s$ [NLO] & $E_T^{\rm miss}$ constrtaint\\
     \hline\hline
     \multirow{6}{*}{\bf S1}
       & Total             & 100 \%  &$\in [300, 350]$~GeV&
           100 \%&$\in [300, 350]$~GeV \\\cline{2-6}
       & $XX$              & $1.6_{-0.1}^{+0.2}\ \%$ &$\in[300, 350]$~GeV&
          $9.4_{-0.6}^{+0.6}\ \%$ & $\in [250, 300]$~GeV\\\cline{2-6}
       & $XY$              & 100 \%&$\in [300, 350]$~GeV&
          100 \%  &$\in [300, 350]$~GeV\\\cline{2-6}
       & $YY$ [total]      & $91.3_{-8.8}^{+6.2}\ \%$ &$\in [300, 350]$~GeV&
          100 \% &$\in [300, 350]$~GeV\\
       & $YY$ [QCD]        & $63.0_{-17.2}^{+20.0}\ \%$ &$\in [300, 350]$~GeV&
          $88.3_{-7.4}^{+4.8}\ \%$ &$\in [300, 350]$~GeV\\
       & $YY$ [$t$-channel]& $70.8_{-4.6}^{+5.0}\ \%$ &$\in [300, 350]$~GeV&
          $87.2_{-1.4}^{+1.0}\ \%$ &$\in [300, 350]$~GeV\\
     \hline\hline\multirow{6}{*}{\bf S2}
       & Total             &  $75.6_{-10.5}^{+10.1}\ \%$ & $\in[700, 800]$~GeV&
          $97.8_{-1.4}^{+0.9}\ \%$ & $\geq 700$~GeV \\\cline{2-6}
       & $XX$              & $0.7_{-0.6}^{+0.6}\ \%$ &$\in [250, 300]$~GeV&
          $3.6_{-0.6}^{+0.3}\ \%$ & $\geq 900$~GeV\\\cline{2-6}
       & $XY$              & $62.7_{-10.4}^{+12.3}\ \%$ &$\in[500, 600]$~GeV&
          $83.9_{-4.3}^{+2.9}\ \%$ &$\in[700, 800]$~GeV\\\cline{2-6}
       & $YY$ [total]      & $24.0_{-3.1}^{+3.1}\ \%$ &$\geq 900$~GeV&
          $58.1_{-3.1}^{+2.2}\ \%$ &$\geq 900$~GeV\\
       & $YY$ [QCD]        & $10.7_{-2.6}^{+4.4}\ \%$ &$\geq 900$~GeV&
          $17.0_{-2.1}^{+2.1}\ \%$ &$\geq 900$~GeV\\
       & $YY$ [$t$-channel]& $29.6_{-2.6}^{+3.3}\ \%$ &$\geq 900$~GeV&
          $38.9_{-1.8}^{+1.2}\ \%$ &$\geq 900$~GeV
  \end{tabular}
  \caption{\it CL exclusions obtained from \ma\ by recasting the
    ATLAS-EXOT-2016-27 analysis~\cite{Aaboud:2017vwy,Fuks:2018yku,
    atlas_exot_2016_27}. The uncertainties are given as absolute
    quantities and originate from scale variations only. When omitted, the
    result is independent of the scale uncertainties. We also indicate the
    $E_T^{\rm miss}$ requirement defining the most sensitive signal region.
    \label{tab:cls}}
   \begin{tabular}{c c | c c c | c c c}
     Sc. & Process & CL$_s$ [LO] & $N_j$ & $M_{\rm eff}$ threshold&
       CL$_s$ [NLO] & $N_j$ & $M_{\rm eff}$ threshold\\
     \hline\hline
     \multirow{6}{*}{\bf S1}
       & Total             &$99.5_{-2.1}^{+0.4}\ \%$ & $\geq 4$ & $>1.4$~TeV&
         100 \% & $\geq 5$ & $>2$~TeV\\\cline{2-8}
       & $XX$              & $0.6_{-0.6}^{+0.6}\ \%$ & $\geq 5$ & $>1.7$~TeV&
          $3.3_{-0.3}^{+0.1}\ \%$ & $\geq 2$ &$>1.6$ TeV\\\cline{2-8}
       & $XY$              &$89.2_{-4.8}^{+4.5}\ \%$ & $\geq 2$ & $>1.6$~TeV&
          $99.8_{-0.2}^{+0.1}\ \%$ & $\geq 5$ & $>2$~TeV\\\cline{2-8}
       & $YY$ [total]      & $96.0_{-7.6}^{+3.4}\ \%$ &$\geq 4$ & $>1.4$~TeV&
          $97.2_{-2.6}^{+1.4}\ \%$ & $\geq 4$ & $>1.4$~TeV\\
       & $YY$ [QCD]        & $88.7_{-14.5}^{+8.8}\ \%$ & $\geq 4$ & $>1.4$~TeV&
          $93.7_{-5.2}^{+2.7}\ \%$ &$\geq 4$ & $>1.4$~TeV\\
       & $YY$ [$t$-channel]& $35.1_{-2.1}^{+3.4}\ \%$ & $\geq 4$ & $>1.4$~TeV&
          $29.7_{-1.4}^{+0.2}\ \%$ &$\geq 5$ & $>2$~TeV\\
     \hline\hline
     \multirow{6}{*}{\bf S2}
       & Total             &$95.0_{-4.3}^{+3.0}\ \%$ & $\geq 2$ & $>1.6$~TeV&
         100 \% & $\geq 2$ & $>1.6$~TeV\\\cline{2-8}
       & $XX$              & $0.6_{-0.6}^{+0.6}\ \%$ & $\geq 6$ & $>2.2$~TeV&
          $1.0_{-0.2}^{+0.0}\ \%$ & $\geq 3$ &$>1.3$ TeV\\\cline{2-8}
       & $XY$              &$61.7_{-7.0}^{+8.4}\ \%$ & $\geq 2$ & $>1.6$~TeV&
          $83.6_{-3.1}^{+1.5}\ \%$ & $\geq 2$ & $>2$~TeV\\\cline{2-8}
       & $YY$ [total]      & $77.4_{-7.5}^{+7.9}\ \%$ &$\geq 2$ & $>1.6$~TeV&
          $97.8_{-1.1}^{+0.5}\ \%$ & $\geq 2$ & $>1.6$~TeV\\
       & $YY$ [QCD]        & $55.3_{-12.3}^{+12.0}\ \%$ & $\geq 2$ & $>2$~TeV&
          $67.7_{-6.4}^{+4.1}\ \%$ &$\geq 2$ & $>1.6$~TeV\\
       & $YY$ [$t$-channel]& $75.6_{-4.8}^{+4.4}\ \%$ & $\geq 2$ & $>2$~TeV&
          $80.1_{-1.6}^{+0.3}\ \%$ &$\geq 2$ & $>1.6$~TeV
  \end{tabular}
  \caption{\it Same as in table~\ref{tab:cls} but for the
    ATLAS-SUSY-2016-07 analysis~\cite{Aaboud:2017phn,atlas_susy_2016_07}. We
    indicate here the jet multiplicity requirement and the effective mass
    $M_{\rm eff}$ threshold defining the most sensitive signal region.
    \label{tab:cls_2}}
\end{table*}

As visible from the results of section~\ref{sec:mctruth}, most of the monojet
signal arises, in the considered {\bf S1} and
{\bf S2} benchmark scenarios, from the production of heavy mediators (by
pairs or in association with dark matter) that then decay into dark matter and
jets. In the following, we reinterpret the results of typical LHC dark
matter searches probing final states featuring a large amount of missing
transverse energy (carried away by the dark matter particles) and an important
hadronic activity. We recast two of such ATLAS analyses for which
reimplementations within the \ma\ Public Analysis Database (PAD) of recasted LHC
analyses~\cite{Dumont:2014tja}\footnote{See the URL
\url{http://madanalysis.irmp.ucl.ac.be/wiki/PublicAnalysisDatabase}.} exist.
Starting from Monte Carlo simulations of the $XX+XY+YY$ dark matter signal
as described in section~\ref{sec:simu}, we make use of \ma\ to
automatically simulate the response of the ATLAS detector by means of
appropriate tunes of the \del\ programme~\cite{deFavereau:2013fsa}. We then
assess the sensitivity of the considered analyses to the {\bf S1} and {\bf S2}
signals by using the CL$_s$ method~\cite{Read:2002hq}.

We consider two ATLAS analyses of 36.2~fb$^{-1}$ of LHC data targeting the
production of missing energy recoiling against at least one hard jet and a
subleading hadronic activity. We recast the ATLAS-EXOT-2016-27
analysis~\cite{Aaboud:2017vwy,Fuks:2018yku,atlas_exot_2016_27} in which the
selection imposes that the dark matter system is produced together with 2 to 4
extra hard jets with quite stringent kinematic requirements. The analysis
includes an ensemble of signal regions that are distinguished by different
inclusive and exclusive
constraints on the missing transverse energy. In the ATLAS-SUSY-2016-07
analysis~\cite{Aaboud:2017phn,atlas_susy_2016_07}, a larger number of jets $N_j$
is allowed ($N_j \geq 2$) and the properties of those jets are less
constrained. The analysis includes several signal regions that mainly differ by
the minimum number of required jets and a constraint on the effective mass
$M_{\rm eff}$ defined by%
\be
 M_{\rm eff} = E_T^{\rm miss} + \sum_{\rm jets} p_T\ .
\ee%

Our results are presented in tables~\ref{tab:cls} and \ref{tab:cls_2} for the
ATLAS-EXOT-2016-27 and ATLAS-SUSY-2016-07 a\-nalysis, respectively. In each
table, we show the confidence level (CL) exclusion obtained when the analysis
signal regions are populated by all the $XX$, $XY$ and $YY$ contributions to the
signal. The impact of the individual channels is also reported, the $YY$
component being further decomposed into its QCD and $t$-channel part. Our
results include theoretical scale uncertainties, which we have extracted by
propagating the uncertainties on the total cross sections down to the CL$_s$
exclusions that we have computed both at LO and NLO. In our recasting procedure,
we conservatively make use of the most sensitive signal region of each analysis,
to derive the exclusion levels, as the statistical model used by the
ATLAS collaboration for the combination of the various regions is not
publicly available. The definition of these regions is provided in
the tables, that hence include the required $E_T^{\rm miss}$ range for the
ATLAS-EXOT-2016-27 analysis, and the thresholds on $N_j$ and $M_{\rm eff}$
for the ATLAS-SUSY-2016-07 analysis.

It turns out that both the {\bf S1} and {\bf S2} scenarios are excluded at the
95\% CL by both analyses, even after accounting for the
uncertainties on the total rates. However, such a conclusion can only be drawn
when more precise NLO simulations are employed and after summing over the
$XX$, $XY$ and $YY$ contributions. As already detailed in
section~\ref{sec:mctruth}, we have found that dark
matter pair production plays no role in the exclusion.

The associated production of a mediator and a dark matter particle ($XY$)
has the largest impact on the ATLAS-EXOT-2016-27 exclusion, the analysis
excluding the {\bf S1} model by solely using this component of the signal.
This stems from an exclusive region in which%
\be
  300~{\rm GeV} \leq E_T^{\rm miss} < 350~{\rm GeV} \ .
\label{eq:s1killer}\ee%
Such a range corresponds to a phase-space region containing a significant
fraction of the $pp\to XY$ events (see figure~\ref{fig:S1_prop}). The
sensitivity to the {\bf S2}
scenario, featuring a much heavier mediator ($m_Y = 1$~TeV), is found to be
slightly below $2\sigma$ when using NLO simulations. In contrast, LO predictions
lead to too conservative conclusions, with a sensitivity barely reaching the
$1\sigma$ level. The LO results are additionally plagued with large scale
uncertainties. The NLO corrections also affect the shapes, and different signal
regions are the most sensitive ones to the LO and NLO signals,%
\be\bsp
  {\rm LO}: &\ 500~{\rm GeV} \leq E_T^{\rm miss} < 600~{\rm GeV} \ ,\\
  {\rm NLO}:&\ 700~{\rm GeV} \leq E_T^{\rm miss} < 800~{\rm GeV} \ .
\esp\ee%

The $pp\to YY$ production cross section being smaller, the
sensitivity of the ATLAS-EXOT-2016-27 analysis to this channel is expected to
be reduced, although the final-state objects that are typically reconstructed
are significantly harder due to the production of two heavy
mediators. In the {\bf S1} scenario, this effect is irrelevant as the mediator
is light enough ($m_Y = 500$~GeV) to be copiously pair-produced. The subsequent
signal is hence excluded by the same signal region as the one defined in
eq.~\eqref{eq:s1killer}. Such a statement can however only be made after using
NLO simulations (the LO rates being too small to reach a 95\% CL exclusion) and
when including not only the QCD-induced production mode, but also the dark
matter $t$-channel exchange one. In the case of the {\bf S2} scenario, the
signal regions are not populated enough to exclude the model. However,
the yields are sufficiently large for driving an exclusion by considering both
the $YY$ and $XY$ contributions, again provided NLO simulations are used.

We derive similar conclusions from the results obtained by
recasting the ATLAS-SUSY-2016-07 analysis. This analysis, that involves more
complex cuts, better depicts the NLO impact on the shapes of the
differential distributions. The corresponding modifications at the differential
level indeed often lead to consider different most sensitive regions at LO and
NLO.

With the examples worked out in this section, we have demonstrated the
importance of relying on new physics precision simulations including NLO QCD
predictions matched with parton showers. The correspondingly more precisely
known total and differential cross sections allow for more robust conclusions on
the sensitivity to the signal. The differences at the level of the distributions
especially play a significant role in modifying the way in controlling how
the different signal regions of the LHC analyses are populated. Moreover, it
is crucial to consider all the components of a given signal,
as their joint contribution may be sufficient to claim an exclusion,
in contrast to the individual contributions taken separately.

\section{Dark matter observables in $t$-channel models}\label{sec:dmsearches}

\subsection{Generalities}\label{sec:gendm}
The studied
$t$-channel simplified models are very peculiar as far as their dark matter
phenomenology is concerned. While tree-level cross sections can be negligible,
if not zero, NLO corrections or loop-induced processes might set up the stage.
This is the case for any considered model restriction involving Majorana or
scalar dark matter~\cite{Giacchino:2013bta,Giacchino:2014moa,Garny:2014waa,
Garny:2015wea,Colucci:2018vxz,Mohan:2019zrk}, while it is more model dependent
for Dirac dark matter~\cite{Ibarra:2015fqa}. In the following, we focus on the
fermionic dark matter case.
In the early universe, the relic abundance is set by the annihilation of
$\tilde{\chi}\tilde{\chi}$ (Majorana) or $\chi\bar{\chi}$ (Dirac) pairs (see
the left diagram in figure~\ref{fig:relic_diag}), unless
the mediator $\varphi$ and the dark matter are within 20\% in mass ($r \equiv
M_\varphi / M_\chi \lesssim 1.2$). In this case,
coannihilations~\cite{Edsjo:1997bg} should be included as they dominate over a
wide range of the parameter space (see the central and right diagrams in figure~\ref{fig:relic_diag}). Moreover, our analysis does not include Sommerfeld
enhancement effects~\cite{Iengo:2009ni,Cassel:2009wt}, as they
are known not to alter the relic density predictions by more than 15\% and only
affect specific parts of the parameter space~\cite{deSimone:2014pda,
Garny:2015wea,Giacchino:2015hvk,Ibarra:2015nca}.

We first consider, as in section~\ref{sec:nlomatch}, a model restriction in
which Dirac dark matter solely couples to the right-handed up quark
({\tt S3D\_uR}). In this model, both the dark matter spin-independent (SI) and
spin-dependent (SD) elastic scattering cross sections off nucleons feature
sizeable tree-level contributions stemming from $s$-channel mediator
exchanges. When coannihilations are negligible ($r\gtrsim 1.2$), indirect
detection rates stem from $\chi \bar{\chi}$ annihilations into pairs of
right-handed up quarks, which proceeds via $s$-wave $t$-channel mediator
exchanges. The associated ve\-lo\-ci\-ty-averaged cross section is about
$3\!\times\! 10^{-26}~\rm{cm^3/s}$ for large $\lambda_{\sss\varphi}$ values, and
thus in the ballpark of the reach of the Fermi-LAT gamma-ray searches from dwarf
spheroidal galaxies~\cite{Fermi-LAT:2016uux}. For illustrative purposes,
we discuss, in the following, relic density and direct detection
predictions. We refer to refs.~\cite{Ibarra:2015fqa,Carpenter:2016thc,
Hisano:2018bpz} for more comprehensive studies.

The Majorana dark matter restriction ({\tt S3M\_uR}) is similar to a
supersymmetric model with bino-like neutralino dark matter and a right-handed up
squark mediator. In this configuration, predictions for direct and indirect
detection observables are dictated by NLO QCD corrections and loop-induced
processes respectively. The direct detection SI elastic scattering cross section
is negligible at tree level because of the Majorana nature of the dark matter,
for which vectorial currents vanish. NLO QCD contributions at one loop, that
include diagrams involving quarks and scalar mediators, therefore dominate and
drive the scattering of dark matter off the nucleon
constituents~\cite{Drees:1993bu,Djouadi:2000ck,Hisano:2010ct,Gondolo:2013wwa,
Hisano:2015bma,Berlin:2015njh}. The SD elastic scattering cross section is, on
the contrary, dominantly dominated by tree-level contributions,
and can be of the same order as the current experimental sensitivity.

Present day \mbox{$\tilde\chi\tilde\chi\to u\bar u$} annihilations in dense
astrophysical environments are $p$-wave suppressed, as the tree-level $s$-wave
contribution is proportional to the up-quark mass that vanishes in the chiral
limit. There however exist two processes that could make Majorana dark matter
detectable: virtual internal bremsstrahlung (VIB) in which the quark pair is
produced together with a photon emitted by the internal $t$-channel propagator,
and loop-induced annihilations into a photon pair or into a photon and a
$Z$-boson. VIB yields a large correction to the tree-level
annihilation cross section, uplifting the $p$-wave suppression by even a few
orders of magnitude, and provides a sharp spectral feature at the highest end of
the gamma-ray spectrum (see, \eg, refs.~\cite{Barger:2011jg,Toma:2013bka,
Garny:2014waa,Giacchino:2013bta,Giacchino:2014moa,Giacchino:2015hvk,
Garny:2015wea,Ibarra:2015nca,Colucci:2018qml}).

On the other hand, annihilations
into photons have been known since a long time as the smoking gun to detect dark
matter, as they produce monochromatic photons pinpointing the dark matter mass
(see, \eg, refs.~\cite{Bouquet:1989sr,Bergstrom:1989jr,Rudaz:1989ij,
Bergstrom:1997fh,Bern:1997ng,Bertone:2009cb}). Whilst these two processes are of
higher order, the astrophysical background for a sharp gamma-ray spectral
feature is very low. This yields a very good experimental sensitivity and
annihilation cross sections well below the canonical
$10^{-26}~{\rm cm}^3/{\rm s}$ value can be probed for a wide range of dark
matter masses~\cite{Ackermann:2015lka}. Moreover, line searches by the HESS
satellite~\cite{Abramowski:2013ax,Abdalla:2016olq} are sensitive to very
heavy dark matter, with masses of tens of TeV, well above the sensitivity range
of the LHC. This thus exhibits a nice complementarity with colliders.

The Majorana dark matter phenomenology briefly sketched here holds for scalar
dark matter too. NLO processes even become relevant at freeze-out, the
tree-level annihilation cross section being $d$-wave suppres\-sed~\cite{%
Giacchino:2013bta}. Similarly, any $t$-channel dark matter model restriction in
which the dark matter couples only to the third generation
requires to account for QCD corrections already for the relic density
predictions~\cite{Colucci:2018vxz}. For instance, for all restrictions of the
{\tt 3rd} type, loop-induced dark matter annihilations into gluons turn out to
be dominant and set the relic density below the $b$-quark
threshold~\cite{Garny:2018icg}. These corrections are typically
not automatically included in available public software such as \maddm\ and
\micromegas, and must be implemented following, \eg,
refs.~\cite{Giacchino:2013bta,Garny:2018icg}.

\subsection{Analysis setup and validation procedure}

In our dark matter analysis, we impose the relic abundance for dark matter to
match the value measured by the Planck satellite in 2018~\cite{Aghanim:2018eyx}.
The direct detection predictions are confronted with the exclusion bounds at 90\% CL of the
XENON1T~\cite{Aprile:2018dbl} and of PICO-60~\cite{Amole:2017dex} experiments
for the SI and SD cases respectively, and we display projections for
the neutrino floor in our SI scattering results~\cite{Billard:2013qya}.
Loop-induced gamma-ray line predictions are compared with the Fermi-LAT~\cite{%
Ackermann:2015lka} and HESS~\cite{Abramowski:2013ax,Abdalla:2016olq} line
searches from the galactic centre, as recasted in ref.~\cite{Garny:2013ama} for
an Einasto dark matter density profile~\cite{Einasto:2009zd} at 95\% CL. We
also show the projected sensitivity of the Cherenkov Telescope Array
(CTA)~\cite{Acharya:2013sxa} as obtained in ref.~\cite{Garny:2013ama} at 95\% CL.

\begin{figure}
  \centering
  \includegraphics[width=.30\columnwidth]{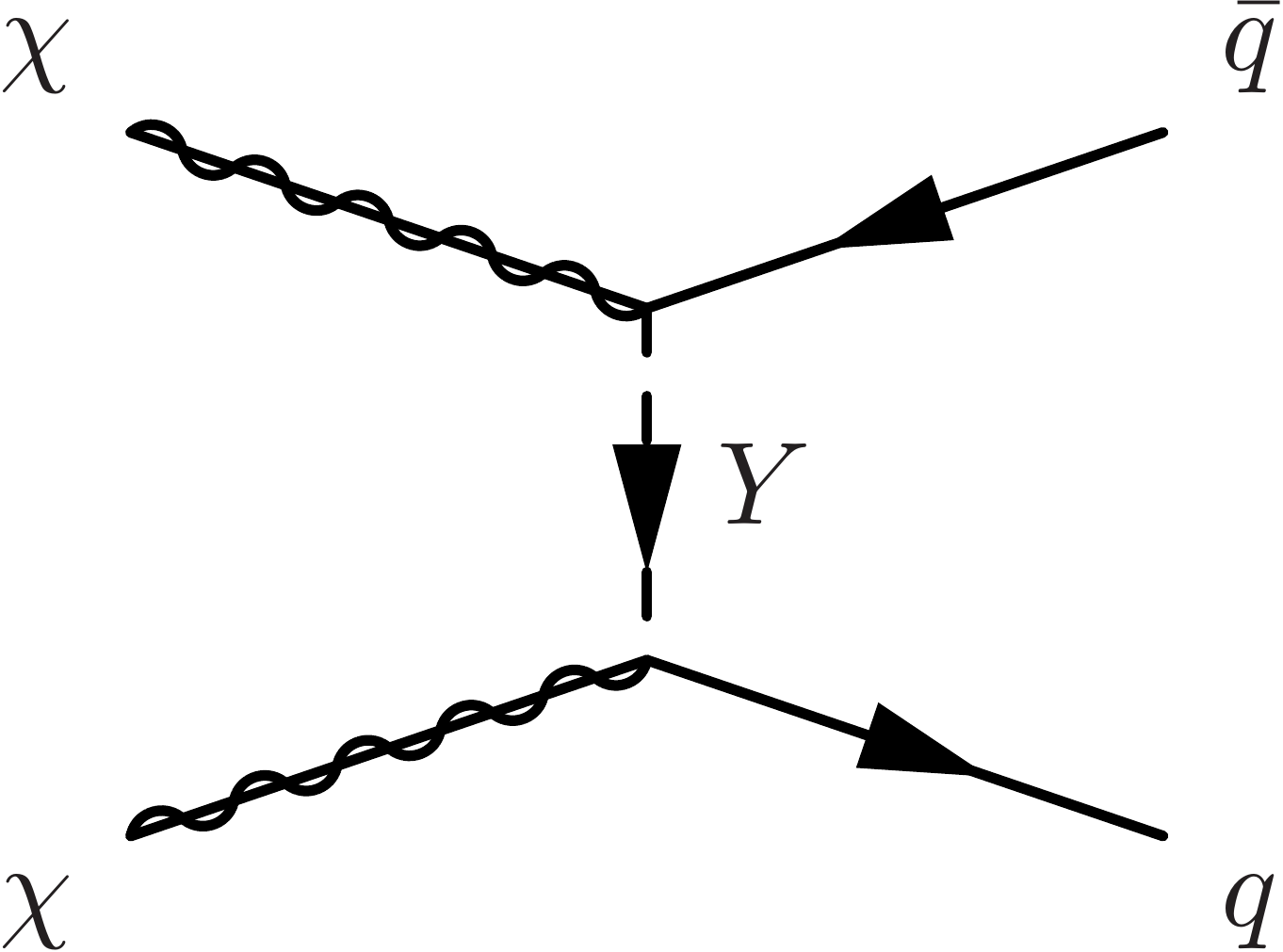}\quad
  \includegraphics[width=.30\columnwidth]{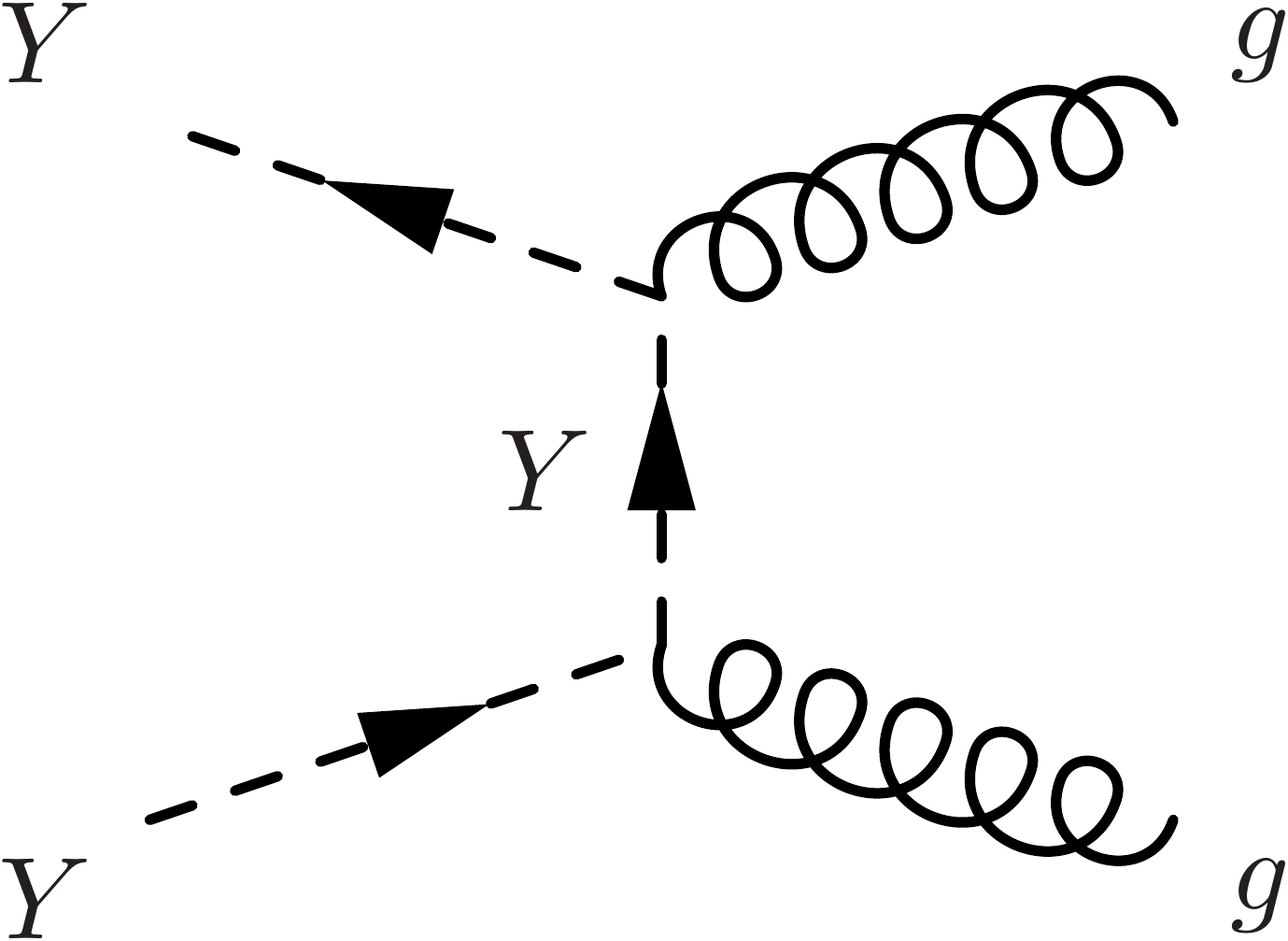}\quad
  \includegraphics[width=.30\columnwidth]{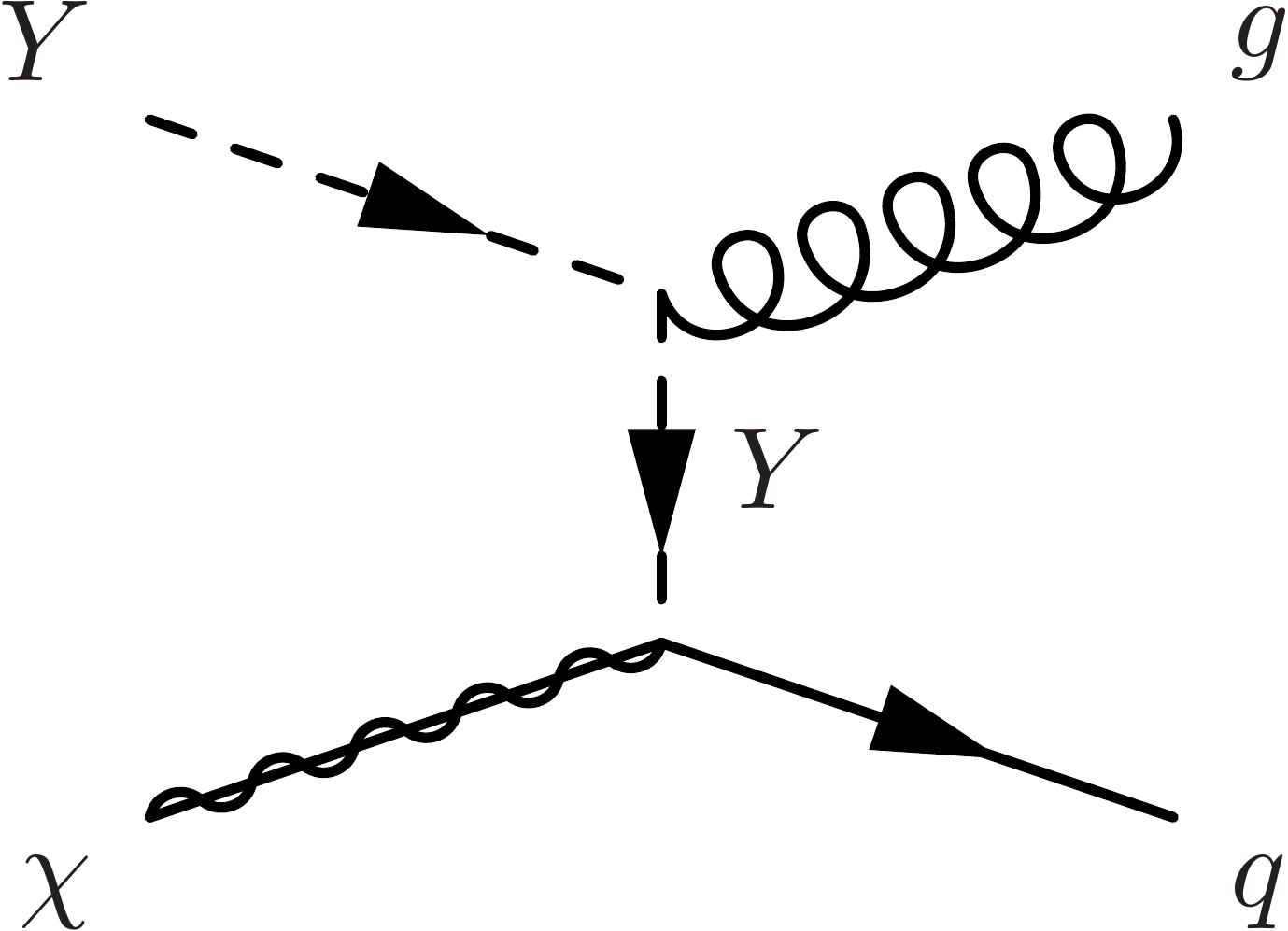}
  \caption{\it Representative LO Feynman diagrams entering the relic density
  computation. We consider dark matter annihilations into quarks (left), as well
  as mediator annihilations (centre) and coannihilations (right) that are
  relevant for mediator and dark matter mass splittings of about 10--20\%.
  \label{fig:relic_diag}}
\end{figure}

In order to compute the relevant observables with \maddm\ in the {\tt
S3D\_uR} model restriction, we type in the command line interface of the
programme,%
\begin{verbatim}
 import model DMSimp_t-S3D_uR --modelname
 define darkmatter xd
 define coannihilator ys3u1
 generate relic_density
 add direct_detection
 output my_project
 launch
\end{verbatim}%
In the case of Majorana dark matter, the model name should be changed to
{\tt DMSimp\_t-S3M\_uR}, and the dark matter candidate name to {\tt xm} (see
table~\ref{tab:fld}). Scans can be performed by using standard \maddm\
syntax~\cite{Ambrogi:2018jqj}, and details for indirect detection calculations
are provided in the next subsections. Representative Feynman
diagrams contributing to the thermally-averaged annihilation cross section
$\langle\sigma v\rangle_{\rm fo}$, assuming a standard dark matter freeze-out
(fo), are depicted in figure~\ref{fig:relic_diag}.

To achieve our calculations in \maddm, we produce a LO UFO library in
which all quarks are massive (unless stated otherwise). It differs from the NLO
UFO library described in section~\ref{sec:general} in which all quarks except
the top quark are massless. We moreover have additionally generated \ch\ model
files~\cite{Belyaev:2012qa}, which is necessary to validate \maddm\ predictions
obtained with the \dmsimpt\ model against known results~\cite{Giacchino:2013bta,
Giacchino:2014moa,Garny:2014waa} derived with \micromegas. All model files are
available from the URL
\url{http://feynrules.irmp.ucl.ac.be/wiki/DMsimpt}.

\subsection{Dark matter observables in the {\tt S3D\_uR} model}
\label{sec:vals3d}

\begin{figure}
  \centering
  \includegraphics[width=\columnwidth]{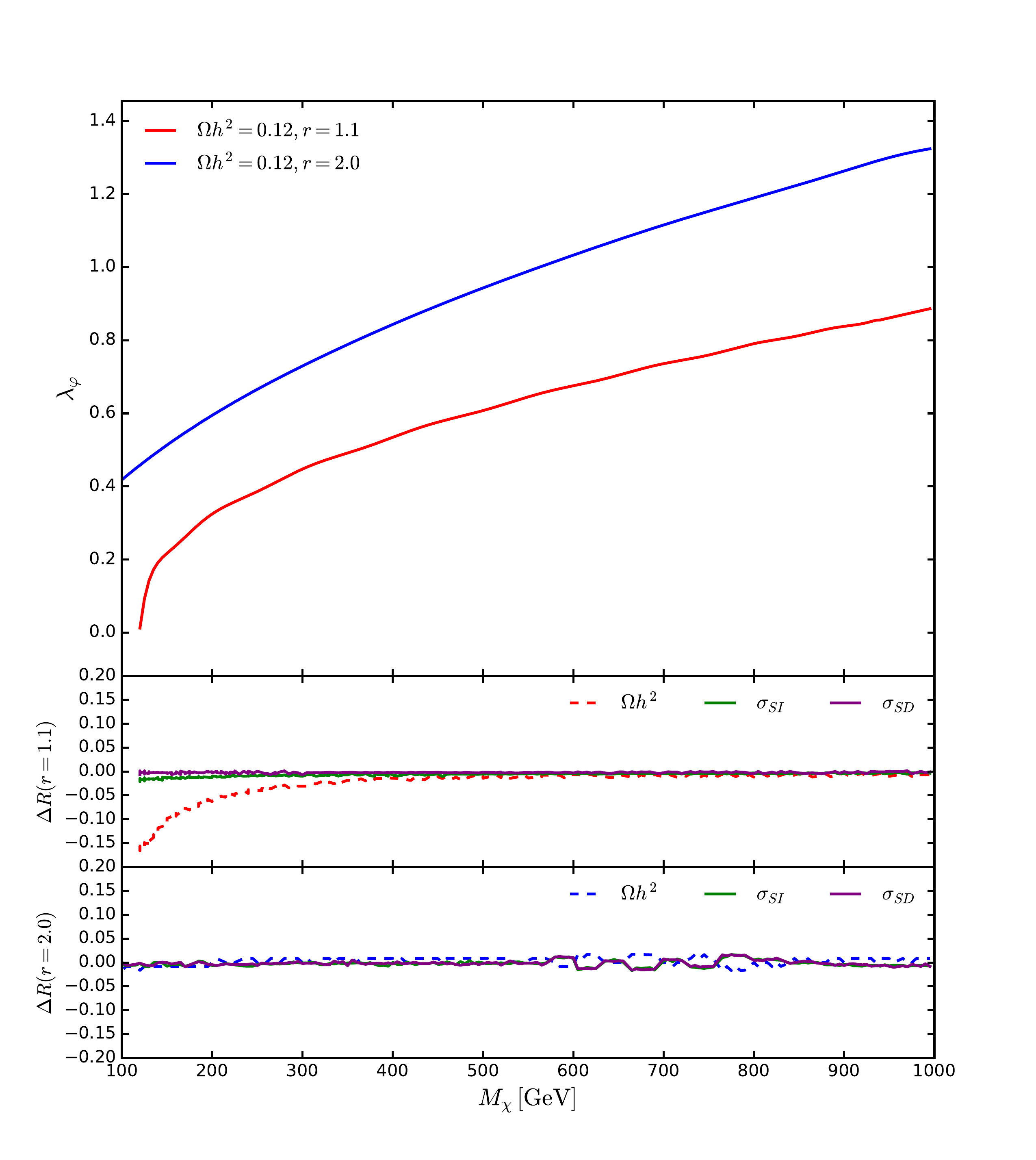}
  \caption{\it Value of the $\lambda{\sss\varphi}$ coupling, as a function of
  the dark matter mass $M_\chi$, leading to a relic density of $\Omega h^2=0.12$
  in the {\tt S3D\_uR} model (top). We present \micromegas\ predictions
  for $r\!=\!1.1$ (red) and 2 (blue). We moreover display the relative
  difference between \micromegas\ and \maddm\ for the relic density
  (dashed), the SI (solid green) and SD (solid magenta) scattering cross section
  off protons for $r\!=\!1.1$ (centre) and 2 (bottom).
  \label{fig:S3D_relic}}
\end{figure}

In figure~\ref{fig:S3D_relic}, we validate our {\tt S3D\_uR} model
implementation by numerically comparing relic density and direct detection
predictions obtained with \maddm\ and \micromegas. In the upper panel, we derive
the $\lambda_{\sss\varphi}$ coupling value required to obtain the correct relic
density as a function of the dark matter mass, for two choices of the mediator
and dark matter mass ratio $r$.

In the more compressed scenario with $r\!=\!1.1$
(red curve), coannihilations and mediator annihilations are important,
especially for small $M_\chi$. For $M_\chi \lesssim 200$~GeV, the relic density
is indeed mostly independent of $\lambda_{\sss\varphi}$, $\langle\sigma v
\rangle_{\rm fo}$ being driven by pure QCD processes involving pairs of
mediators annihilating into quarks and gluons (second diagram in
figure~\ref{fig:relic_diag}). To properly evaluate these QCD processes, we
include the running of the strong
coupling in \maddm\footnote{This feature is now standard in the publicly
available version of \maddm\ at the URL \url{https://launchpad.net/maddm}.},
as implemented by default in \mg~\cite{Campbell_1999}. The number of quarks
included in the loops depends on the running scale (and can be at most 5),
and the QCD beta function can be evaluated at 1, 2 (default) and 3 loops.
As far as the dark matter mass increases, processes involving both $\chi$ and
$\varphi$ become relevant so that $\lambda_{\sss\varphi}$ has to be sizeable to
obtain the right relic density. Already for $M_\chi\sim250$ GeV, $XX$
annihilations (first diagram in figure~\ref{fig:relic_diag}) and $XY$
coannihilations (third diagram in figure~\ref{fig:relic_diag}) contribute to the
total scattering cross section $\langle\sigma v\rangle_{\rm fo}$ by about 45\%
and 30\% respectively, the reminder being due to mediator-pair annihilations
($YY$). As in the previous section, $X\!=\!\chi,\bar\chi$ and $Y\!=\!\varphi,%
\varphi^\dag$ in our notations.
For dark matter masses larger than 500~GeV, $XX$ annihilations
dominate, the $XY$ and $YY$ processes contributing only to less than about 20\%
to the relic density.

In the $r\!=\!2$ case (blue line), $Y$ is too heavy
relatively to dark matter to be relevant at freeze-out. Only $XX$
annihilations contribute, and $\lambda_{\sss\varphi}$ has to be sizeable and
larger than for $r\!=\!1.1$ for any a given $M_\chi$ value. This large value
compensates the smaller $\langle\sigma v
\rangle_{\rm fo}$ cross section stemming from a smaller number of relevant
processes than in the $r\!=\!1.1$ scenario where coannihilations and mediator
annihilations play a role.

In order to quantify the numerical differences between \micromegas\ and \maddm\
predictions for a given dark matter observable $\mathcal{O}$, we define the
quantity%
\be
  \Delta R = \frac{\mathcal{O}_{\text{\micromegas}}-\mathcal{O}_{\text{\maddm}}}
  {\mathcal{O}_{\text{\micromegas}}}\ .
\ee%
and focus, in figure~\ref{fig:S3D_relic}, on the relic density (dashed), and the
SI (solid green) and SD (solid magenta) direct detection cross sections. We
present the dependence of $\Delta R$ on the dark matter mass both for the
$r\!=\!1.1$ (middle panel) and $r\!=\!2$ (lower panel) scenarios. Predictions
for both the SI and SD dark matter scattering cross section off
protons\footnote{Similar results are obtained in the neutron case.} are
found in perfect agreement, the discrepancy between \maddm\ and \micromegas\
being of at most a few percents for the probed dark matter mass range.
Relic density predictions are also found to agree quite well, except for dark
matter masses in the 100--200 GeV range for the $r=1.1$ configuration. In this
parameter space region, we get a discrepancy reaching 5\% to 15\% due to the
different treatment of the QCD sector in both codes (the relic density being
driven by QCD-induced mediator annihilations). \micromegas\ indeed includes
running quark masses, in addition to the strong coupling
running~\cite{Belanger:2014vza}.

\subsection{Dark matter observables in the {\tt S3M\_uR} model}

\begin{figure}
  \centering
  \includegraphics[width=\columnwidth]{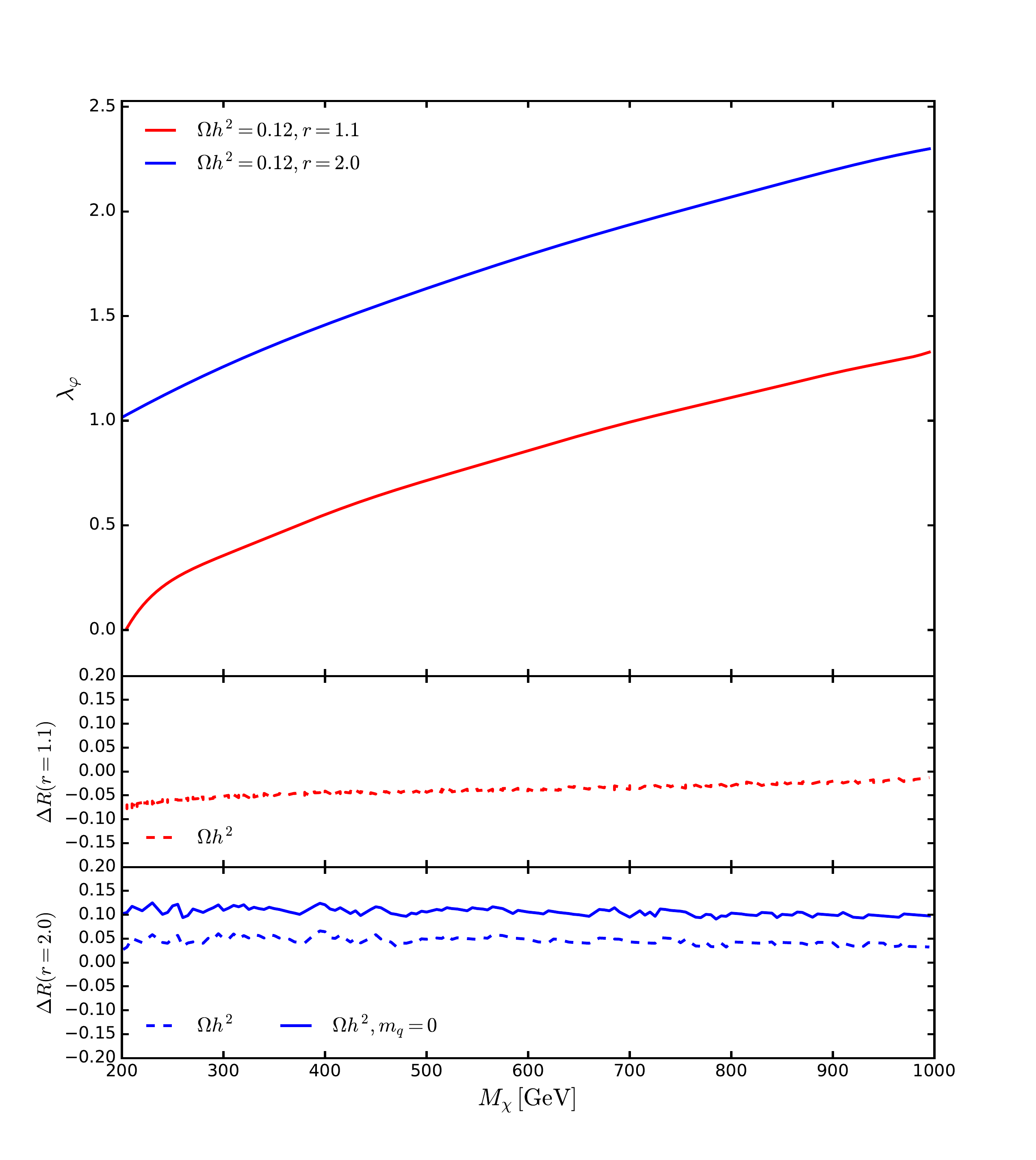}
  \caption{\it Value of the $\lambda{\sss\varphi}$ coupling, as a function of
  the dark matter mass $M_\chi$, leading to a relic density of $\Omega h^2=0.12$
  in the {\tt S3M\_uR} model (top). We present \micromegas\ predictions
  for $r\!=\!1.1$ (red) and 2 (blue). We moreover display the relative
  difference between \micromegas\ and \maddm\ for the relic density
  (dashed) for $r\!=\!1.1$ (centre) and 2 (bottom). In the last case,
  the limits in which all quarks are massless are also presented (solid).
  \label{fig:S3M_relic}}
\end{figure}

In this section, we focus on Majorana dark matter and derive, in figure~\ref{%
fig:S3M_relic}, the values of the $\lambda_{\sss\varphi}$ coupling that are
needed to obtain the correct relic density for $r\!=\!1.1$ (red) and $r\!=\!2$
(blue) configurations. Comparing with the Dirac dark matter case, larger
couplings are generally required as a consequence of the Majorana nature of
dark matter, with the exception of setups featuring dark matter masses below
500~GeV where the relic density is driven by mediator annihilations and
coannihilations. Another remarkable difference with Dirac dark matter, in the
$r\!=\!1.1$ scenario, is that there is no phenomenologically-viable solution
for $M_\chi \lesssim 200$~GeV. In the middle and lower panel of the figure, we
show that \maddm\ and \micromegas\ predictions agree quite well, as
$\Delta R \lesssim 5\%$ for both scenarios (dashed lines), except for light dark
matter where more important differences stem from the different treatment of the
QCD sector. We additionally assess the impact of the quark masses that induce a
10\% shift (including running quark mass effects) relatively to the values
obtained with \micromegas\ (for a massive quark setup).

\begin{figure}
  \centering
  \includegraphics[width=.95\columnwidth]{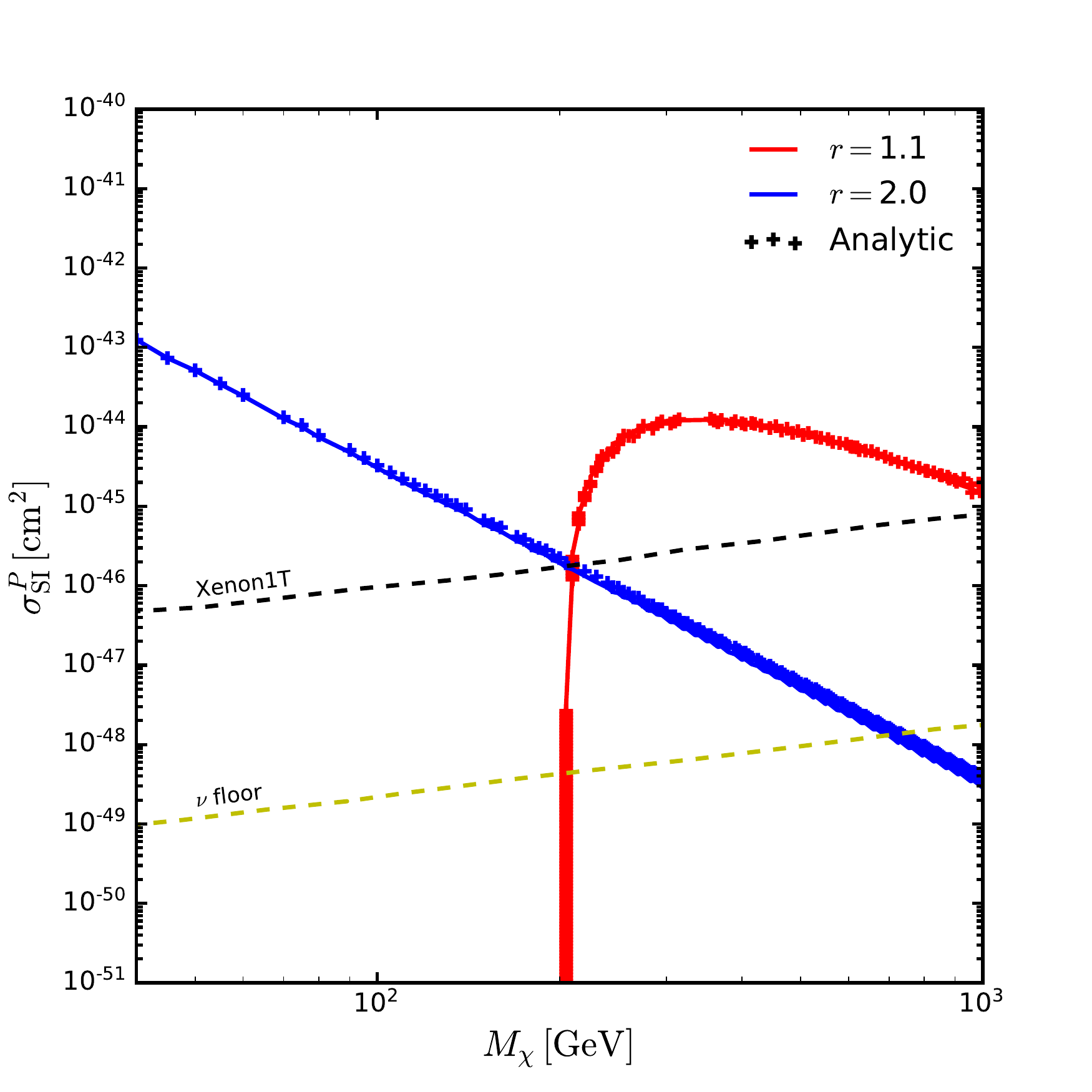}
  \caption{\it NLO SI scattering cross section off protons as a
  function of the dark matter mass for $r\!=\!1.1$ (red) and $r\!=\!2$
  (blue), with a $\lambda_{\sss\varphi}$ coupling yielding the right relic
  density. We compare \micromegas\ predictions (solid lines) with the analytical
  results of ref.~\cite{Hisano:2015bma}. We additionally include the exclusion
  bounds extracted from current XENON1T results~\cite{Aprile:2018dbl} (dashed
  black line) and the neutrino floor~\cite{Billard:2013qya} (dashed yellow
  line).\label{fig:S3M_dd}}
  \includegraphics[width=.95\columnwidth]{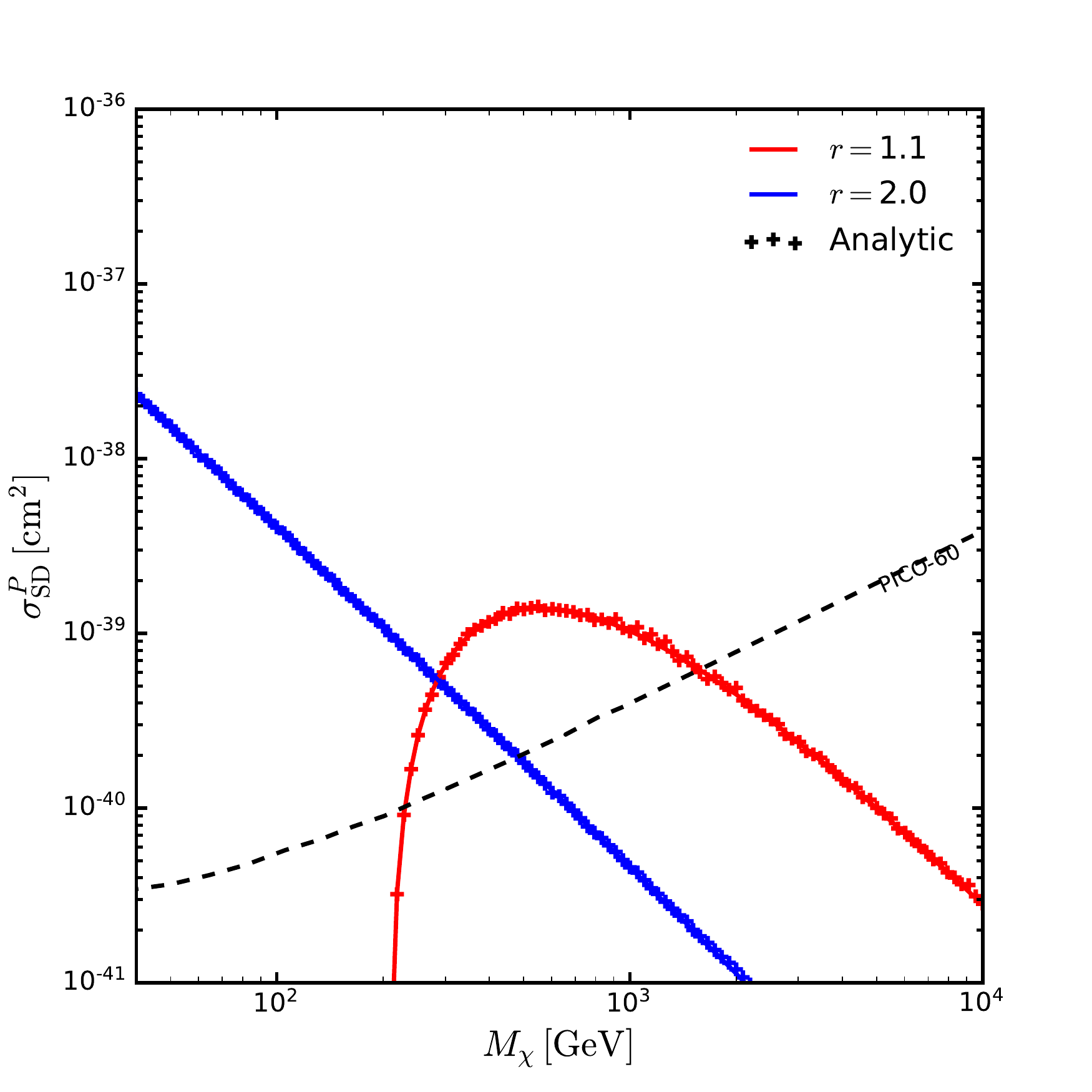}
  \caption{\it Same as figure~\ref{fig:S3M_dd} but for the SD scattering cross
  section. The predictions are compared with the current PICO-60 exclusion
  bounds~\cite{Amole:2017dex} (dashed black line). \label{fig:S3M_dd2}}
\end{figure}

In figure~\ref{fig:S3M_dd}, we estimate the NLO SI dark matter elastic
scattering cross section off protons for the $r\!=\!1.1$ (red) and $r\!=\!2$
(blue) scenarios. For each dark matter mass value, we fix the
$\lambda_{\sss\varphi}$ coupling to reproduce the relic density as observed by
the Planck collaboration. We compare the NLO predictions obtained with \micromegas\
(solid curves) with the total SI cross-section (markers) given by the sum of the LO contribution obtained with \maddm\ with the
analytically available NLO corrections from ref.~\cite{Hisano:2015bma}. The
analytic expression used here read
\be\bsp
\sigma^p_{\rm SI} &= \frac{4}{\pi} \frac{M_p^2 M_\chi^2}{(M_p + M_\chi)^2} f_p^2\ , \\
\sigma^p_{\rm SD} &= \frac{12}{\pi} \frac{M_p^2 M_\chi^2}{(M_p + M_\chi)^2} a_p^2\ ,
\esp\ee
where $M_p$ is the proton mass and the form factors $f_p$ and $a_p$ are
functions of the Wilson coefficients describing the effective interactions with
the proton components. These form factors are listed, for the various models, in
ref.~\cite{Hisano:2015bma}.
In the case of our {\tt S3M\_uR} model, the SD form factor is given by
\be\bsp
a_p = \frac{1}{8}\frac{1}{M_\varphi^2 - M_\chi^2} \Delta u_p
\esp\ee
where $\Delta u_p=0.842$ is the spin fraction of the up-quark in the proton.

Concerning \micromegas\, the NLO SI contribution is automatically included following ref.~\cite{Drees:1993bu}.
An excellent agreement is found. 
Confronting those results to the exclusion limits of XE\-NON~1T~\cite{Aprile:2018dbl}, half of the viable parameter space ($M_\chi
\lesssim 150$~GeV) is excluded for both spectrum compression options. Moreover,
this shows that most of the currently viable parameter space can be explored by
next-generation
dark matter experiments, as the corresponding SI scattering cross sections are
larger than the expectation of the neutrino background~\cite{Billard:2013qya}.

Predictions for the LO SD elastic dark matter cross section off protons are
shown in figure~\ref{fig:S3M_dd2}, in which we demonstrate the agreement between
the numerical results of \maddm\ (solid lines) and the analytic expressions of
ref.~\cite{Hisano:2015bma} (markers).
Confronting those
predictions with the exclusion limits obtained from the PICO-60
experiment~\cite{Amole:2017dex}, it turns out that dark matter masses smaller
than 500~GeV in the $r\!=\!2$ configuration, and lying in the [250, 1200]~GeV
range in the $r\!=\!1.1$ case, are disfavoured. Whereas the running of the
$\lambda_{\sss\varphi}$ coupling from the electroweak scale down to the GeV
scale is known to largely enhance direct detection cross section
predictions~\cite{Mohan:2019zrk}, this effect is not included neither in \maddm\
nor in \micromegas. This goes beyond the scope of this work.

\begin{figure}
  \centering
  \includegraphics[width=\columnwidth]{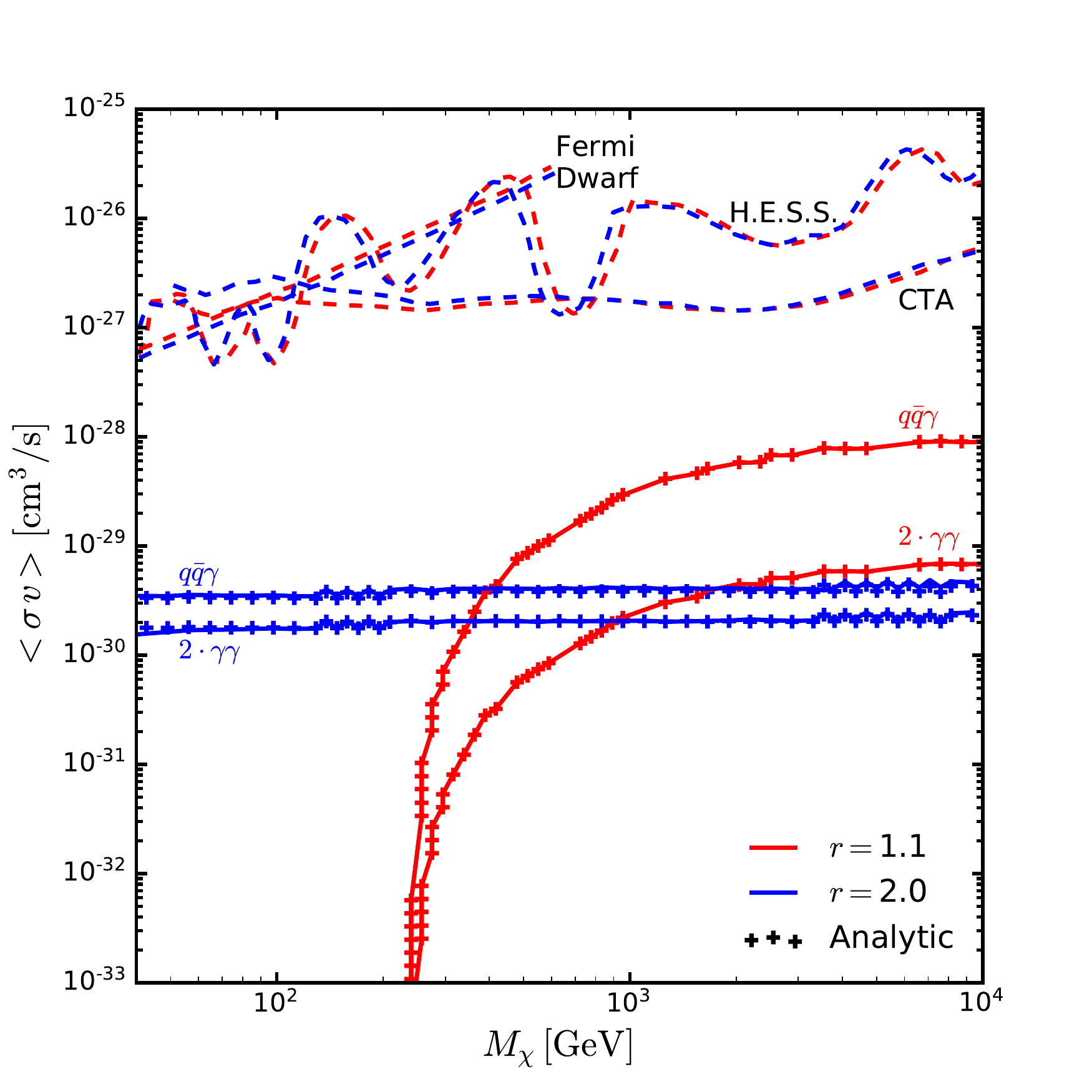}
  \caption{\it Present time dark matter annihilation cross section as a function
  of $M_\chi$ in the $r=1.1$ (red) and $r=2$ (blue) configurations.
  Predictions for the VIB ($q\bar q\gamma$) and photon-pair production
  ($\gamma\gamma$) modes have been automatically computed with \maddm, and
  compared with the analytic expressions of refs.~\cite{Giacchino:2014moa} and
  \cite{Giacchino:2013bta} for the VIB and $\gamma\gamma$ processes
  respectively. We additionally show constraints (dashed) from the
  Fermi-LAT dwarf spheroidal galaxies measurements and from the HESS
  experiment~\cite{Garny:2013ama}, as well as the expected sensitivity of the
  CTA experiment~\cite{Garny:2013ama}. \label{fig:S3M_id}}
\end{figure}

We finally consider dark matter indirect detection in figure~\ref{fig:S3M_id},
in which we present predictions for the present time dark matter annihilation
cross section in the $\tilde\chi\tilde\chi \to u\bar{u}\gamma$ and $\tilde\chi
\tilde\chi\to\gamma\gamma$ channels for both considered benchmark scenarios. In
both cases, the VIB cross section is larger than the diphoton one, although the
latter loop-induced rate is of a similar order of magnitude as the former
three-body one\footnote{The
extra factor of two accounts for the photon multiplicity.}. For more split
spectra, or equivalently for larger $r$ values, the loop-induced contributions
are however known to dominate~\cite{Garny:2014waa}. Confronting our
predictions with the experimental results that are very sensitive to sharp
features and lines in the gamma-ray spectra, we observe that there is no
sensitivity to the two
considered benchmark scenarios. This holds both for current exclusions extracted
from the Fermi-LAT dwarf spheroidal galaxy measurements or HESS data, and for
the expectation of the future CTA line search from the galactic centre.

In our predictions, we have compared the results of \maddm\
obtained by using the NLO \dmsimpt\ UFO library of section~\ref{sec:general}
(solid lines) with the analytical approximated expressions of
refs.~\cite{Giacchino:2014moa} and \cite{Giacchino:2013bta} for the VIB and
diphoton channels, respectively, 
\be\bsp
  \langle \sigma v \rangle_{\gamma \gamma} &= \left(\frac{4}{3}\right)^2
    \frac{\alpha_{\rm EM}^2 \lambda_\varphi^4}{256\pi^3M_\chi^3} I(r)\ ,\\
  \langle \sigma v \rangle_{q \bar{q} \gamma} &=
    \frac{\alpha_{\rm EM} \lambda_\varphi^4}{48 \pi^2 M_\chi^2}F(r) \ ,
\esp\ee
where $\alpha_{\rm EM}$ denotes the electromagnetic coupling constant
and
\be\bsp
  I(r) &= \int_0^1\frac{{\rm d}x}{x}\log\left|\frac{-x^2 + (1 - r^2)x + r^2}{x^2 + (-1-r^2)x + r^2}\right| \ ,\\
  F(r) &= (r^2\!+\!1)\left(\frac{\pi^2}{6} - \log^2\frac{r^2+1}{2r^2}\right)
    - 2{\rm Li}_2\frac{r^2+1}{2r^2}\\ &\quad
   + \frac{4r^2 + 3}{r^2 +1} + \frac{4r^2 - 3r^2 -1}{2r^2}\log\frac{r^2-1}{r^2+1} \ .
\esp\ee%
As discussed in the
manual~\cite{Ambrogi:2018jqj}, \maddm\ can automatically handle $2\to3$ generic
annihilation processes by typing in,%
\begin{verbatim}
 import model DMSimp_t-S3M_uR --modelname
 define darkmatter xm
 define coannihilator ys3u1
 generate indirect_detection u u~ a
 output my_project
 launch
\end{verbatim}%
The results of figure~\ref{fig:S3M_id} however represents the first validation
of a fully automated loop-induced process computation for a dark matter
observable\footnote{ The wiggles in both the numerical and analytical estimations in figure~\ref{fig:S3M_id} are numerical artefacts which can be smoothed out by performing a higher resolution scan.}. This feature will be available from the future version of
\maddm\footnote{This release can already be obtained from the authors.}.

\section{Summary and conclusion}\label{sec:concl}
In this work, we have introduced the \dmsimpt\ framework for dark matter
$t$-channel models, available from the URL
\url{http://feynrules.irmp.ucl.ac.be/wiki/DMsimpt}.
This consists in a unique \fr\ implementation that allows
for the calculation, through the various high-energy physics tools interfaced
to \fr, of a large set of dark matter observables at colliders and in cosmology.
The model is shipped with several restrictions relevant for simplified models
featuring dark matter and coloured mediators of different spins.
We have extensively studied two of those restrictions in which the dark matter
is either a Dirac or a Majorana fermion, and the mediator is a scalar
state coupling to the right-handed up quark.

We have generated a UFO model including ingredients for the automatic
computation of collider observables matching NLO QCD predictions with parton
showers. By a joint use of the \mg, \py, \ma, \fj\ and \del\ program\-mes, we
have investigated the impact of the NLO corrections on various
observables relevant for typical dark matter searches at the LHC through monojet
pro\-bes, and shown how this could affect the sensitivity of the corresponding
experimental searches. Our results emphasise the benefits of using NLO
simulations to get more realistic predictions for total and differential cross
sections including smaller theoretical systematics. We have
moreover demonstrated how considering all new physics signals predicted by a
given scenario as a whole is necessary for a better assessment of the LHC
sensitivity to new phenomena. At the NLO accuracy, this however requires a
specific treatment of $s$-channel resonant contributions usually appearing in
the real emission contributions in order to avoid their double counting. Such a
task can be automatically achieved within the \mg\ framework.

We have then made use of the \maddm\ programme for the automatic calculation of
the dark matter relic density, spin-independent and spin-dependent
scattering cross sections off nucleons and indirect detection rates. We have
validated our predictions through a comparison with \micromegas\ (using a \ch\
model file generated from our general \dmsimpt\ \fr\ implementation) and
existing analytical calculations. Our predictions include both NLO
corrections and the contributions of loop-induced processes as they could
be dominant in specific model configurations, in particular for what concerns
Majorana dark matter spin-independent direct detection (that is strongly
affected by higher orders) and indirect detection (driven by loop-induced and
virtual internal bremsstrahlung subprocesses). While \micromegas\ can account
for corrections to direct detection, \maddm\ can automatically
evaluate VIB processes. We have moreover pioneered the first
automatic calculation of a loop-induced contribution to the production of
gamma-ray lines by dark matter annihilations at present time.

In conclusion, our work presents, for the first time, a unified framework
to undertake precision dark matter calculations in cosmology and at colliders
for a large class of $t$-channel dark matter models.

\begin{acknowledgements}
The authors are grateful to L. Lopez Honorez, K. Mohan, A. Pukhov and T. Tait
for their help and useful comments during the model validation steps.
CA is supported by the Innoviris  ATTRACT 2018 104 BECAP 2 agreement.
This work has received funding from the European Union's Horizon 2020 research and innovation programme as part of the 
Marie Skłodowska-Curie Innovative Training Network MCnetITN3 (grant agreement no. 722104).
\end{acknowledgements}

\bibliographystyle{jhep}
\bibliography{library}
\end{document}